\documentclass[aps,pra,reprint,groupedaddress]{revtex4-2}
\usepackage{graphicx}										% include figure files
\usepackage{dcolumn}										% Align table columns on decimal point
\usepackage{color}											% Allows for colored text
\usepackage{hyperref}										% Allows for hyperlinked references.
	\hypersetup{
    	colorlinks=true,     								% Colors text in linked references, instead of boxing
    	urlcolor=blue,										% Makes URL links blue
	}

\begin{document}

\title{Doubly-ionized lanthanum as a qubit candidate for quantum networks}

\author{S. Olmschenk}
	\email{steven.olmschenk@denison.edu}
\affiliation{Department of Physics and Astronomy, Denison University, Granville, Ohio 43023, USA}

\date{\today}

%%%%%%%%%%%%%%%%%%%%%%%%%%%%%%%%%%%%%%%%%%
\begin{abstract}
We propose doubly-ionized lanthanum (La$^{2+}$) as a possible qubit candidate for quantum networks.  Transitions between the lowest levels in the atom are in the infrared, enabling a direct matter-light interface amenable to long-distance quantum communication.  These transitions could also be used to directly laser-cool trapped La$^{2+}$ ions.  The rich hyperfine structure of the ion may allow for a qubit stored in magnetic-field insensitive states, as well as protocols for atom-photon entanglement.  
\end{abstract}

\maketitle

%%%%%%%%%%%%%%%%%%%%%%%%%%%%%%%%%%%%%%%%%%
\section{Introduction}
	\label{sec:intro}
Quantum information has the potential to revolutionize computation and communication.  Trapped atomic ions are a leading platform for applications in quantum information, due in part to their long coherence times~\cite{langer2005:long-lived_qubit, harty2014:high-fidelity_ion_qubits, wang2017:qubit_coherence, wang2021:qubit_coherence}, and precise control and detection of the atomic quantum states using microwave and laser radiation~\cite{harty2014:high-fidelity_ion_qubits, gaebler2016:high-fidelity_gate_set_Be_ions, christensen2020:high-fidelity_qubit, zhukas2021:detection_qubit_register}.  These pristine quantum systems continue to extend the frontiers of quantum computation~\cite{blatt2008:ions_entangled, bruzewicz2019:trapped_ion_qc_review, alexeev2021:q_comp_sci_discovery, brown2021:materials_challenges_qc, deleon2021:materials_challenges_qc}, quantum simulation~\cite{blatt2012:quantum_simulation, monroe2021:programmable_quantum_simulations, altman2021:q_sim_roadmap}, and quantum sensing~\cite{ludlow2015:optical_atomic_clocks, degen2017:quantum_sensing}. 

Trapped atomic ions are also a candidate for quantum networks~\cite{briegel1998:quantum_repeater, simon2003:dist_entangle, kimble2008:qinternet, sangouard2009:q-repeater_ions, duan2010:quantum_networks_ions, awschalom2021:quantum_interconnects}.  Pioneering experiments have demonstrated the ability to entangle ions remotely, over considerable distance, using their emitted photons~\cite{moehring:ion-ion, olmschenk:teleportation, hucul2015:modular_entanglement,  stephenson2020:remote_entanglement}.  In general, though, the ultraviolet and visible wavelengths of photons directly produced by strong transitions between low-lying levels in atomic ions~\cite{bruzewicz2019:trapped_ion_qc_review} are not conducive to long-distance transmission through optical fibers~\cite{winzer2018:fiber-optic_transmission, sibley2020:optical_comm}.  Nevertheless, recent experiments demonstrated the ability to convert the short-wavelength photons emitted by trapped ions to the infrared frequencies more amenable to long-distance communication~\cite{mcguinness2010:freq_translation, zaske2012:freq_conversion, kim2013:freq_conversion, kasture2016:freq_conversion, kambs2016:freq_conversion, siverns2017:freq_conversion, rutz2017:freq_conversion, bock2018:freq_conversion, meraner2020:ion_network_node, hannegan2021:freq_conv}.  These experiments represent a promising approach to quantum networks with trapped ions, even though the frequency-conversion stage adds noise and loss, as well as additional complexity to an already intricate system.

Here, doubly-ionized lanthanum (La$^{2+}$) is proposed as an alternative approach to quantum networks.  Transitions between the lowest-lying levels in La$^{2+}$ are in the infrared, with low-loss transmission in standard optical fiber, and may be used for laser cooling.  The rich hyperfine structure of the ion allow for magnetic-field insensitive qubit states, as well as state detection fidelities comparable to other atomic ions, and methods for directly producing ion-photon entanglement at wavelengths amenable to long-distance transmission for the realization of quantum networks.

%%%%%%%%%%%%%%%%%%%%%%%%%%%%%%%%%%%%%%%%%%
\section{Atomic structure and laser cooling}
	\label{sec:structure_cooling}
\begin{figure*}
  \includegraphics[width=2.0\columnwidth,keepaspectratio]{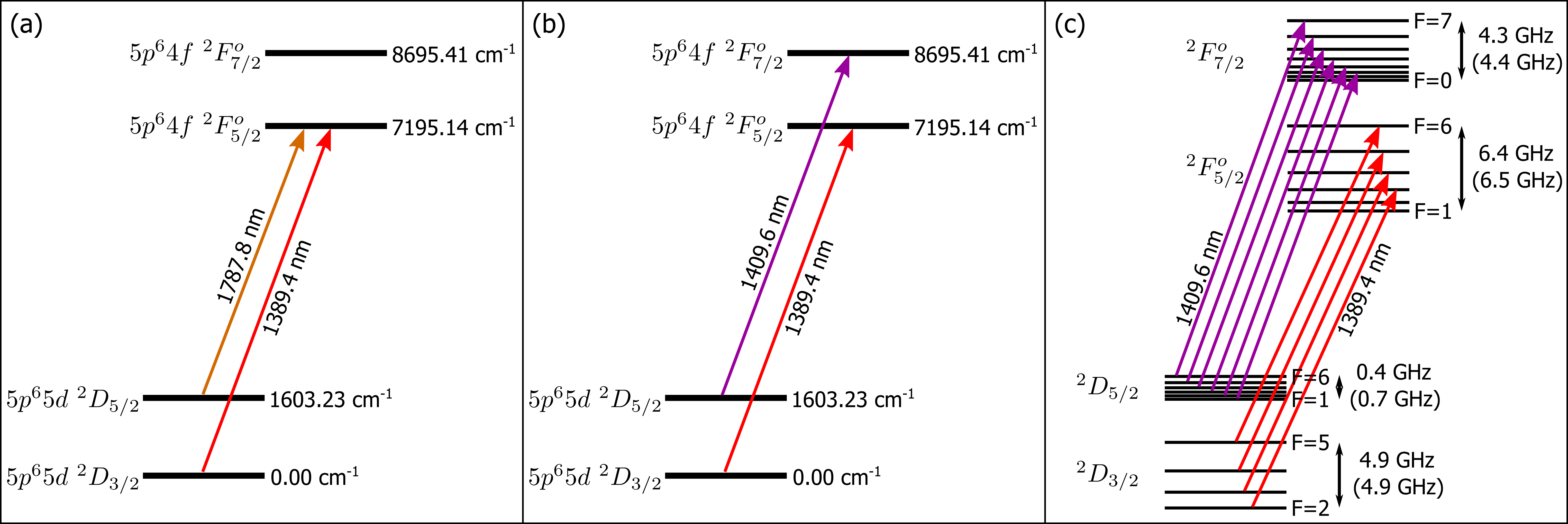}
\caption{Energy level diagram of La$^{2+}$ with possible laser cooling schemes.  (a) A 3-level cooling scheme driving the ${}^2D_{3/2} \leftrightarrow {}^2F^o_{5/2}$ transition near 1389.4 nm and the ${}^2D_{3/2} \leftrightarrow {}^2F^o_{5/2}$ transition near 1787.8 nm.  (b) A 4-level cooling scheme driving the ${}^2D_{3/2} \leftrightarrow {}^2F^o_{5/2}$ transition near 1389.4 nm and the ${}^2D_{5/2} \leftrightarrow {}^2F^o_{7/2}$ transition near 1409.6 nm.  (c) The 4-level cooling scheme with hyperfine structure shown (not to scale), with light driving each ${}^2D$ $| F = n \rangle$ to ${}^2F^o$ $| F = n+1 \rangle$ transition.  Here each top (bottom, parenthetical) frequency gives the extent of the hyperfine manifold splitting calculated from the experimental (theoretical) hyperfine $A$ and $B$ coefficients given in Ref.~\cite{olmschenk2017:laiii_hyperfine} (Ref.~\cite{li2021:hyperfine_laiii}).  Wavelengths given in air~\cite{NIST:ASD_2020}.}
\label{fig:cooling_level_diagram}       % Give a unique label
\end{figure*}

The lowest levels in doubly-ionized lanthanum are ${}^2D_{3/2}$ and ${}^2D_{5/2}$, followed by ${}^2F^o_{5/2}$ and ${}^2F^o_{7/2}$, as shown in Fig.~\ref{fig:cooling_level_diagram}.  The ${}^2F^o_{5/2}$ and ${}^2F^o_{7/2}$ levels are predicted to have lifetimes of about 4.6 $\mu$s and 4.4 $\mu$s, respectively, while the metastable ${}^2D_{5/2}$ is predicted to have a lifetime of about 15 s~\cite{safronova2014:laiii}.  All four low-lying levels have substantial hyperfine structure, as both naturally occurring isotopes have relatively large nuclear spin:  the ${}^{139}$La isotope has a natural abundance of 99.91\% and nuclear spin $I=7/2$; the ${}^{138}$La isotope has a natural abundance of 0.09\% and nuclear spin $I=5$~\cite{delaeter2003:atomic_weights}.  The large nuclear spin combined with the substantial $J$-values of the electronic levels results in a rich hyperfine structure that has recently been measured~\cite{olmschenk2017:laiii_hyperfine} and calculated~\cite{li2021:hyperfine_laiii} for the ${}^{139}$La isotope.  Despite multiple hyperfine manifolds, the ion is tractable for laser cooling.

% Figure

Trapped La$^{2+}$ ions could be laser-cooled by using either a 3-level ($\Lambda$-type) or 4-level scheme.  A possible 3-level cooling scheme involves driving the ${}^2D_{3/2} \leftrightarrow {}^2F^o_{5/2}$ transition near 1389.4 nm (air), along with the ${}^2D_{3/2} \leftrightarrow {}^2F^o_{5/2}$ transition near 1787.8 nm (air), as shown in Fig.~\ref{fig:cooling_level_diagram}(a).  This 3-level cooling scheme would be similar to the approach often used for cooling other trapped ions with a low-lying $D$-level, including Ca${}^+$~\cite{urabe1992:ca_ion_cooling}, Sr${}^+$~\cite{madej1990:sr_ion_laser_cooling}, and Ba${}^+$~\cite{neuhauser1978:ba_ion_cooling}, and is susceptible to coherent population trapping~\cite{gray1978:coherent_pop_trapping}.   Alternatively, a possible 4-level cooling scheme involves driving the ${}^2D_{3/2} \leftrightarrow {}^2F^o_{5/2}$ transition near 1389.4 nm (air) and the ${}^2D_{5/2} \leftrightarrow {}^2F^o_{7/2}$ transition near 1409.6 nm (air), as shown in Fig.~\ref{fig:cooling_level_diagram}(b), which would be similar to the approach often used for cooling Yb${}^+$~\cite{bell:four-level, olmschenk:state-detect}.  In either case, all necessary frequencies for laser cooling can be generated by just two lasers, with wavelengths accessible using semiconductor diode lasers.  

The 4-level cooling scheme has potential advantages.  First, this option has the important technical advantage that the two required laser wavelengths are near enough to each other that a common set of fibers and optics may be used.  Second, in this case there exists a true cycling transition between the ${}^2D_{5/2}$ $| F = 6 \rangle$ and ${}^2F^o_{7/2}$ $| F = 7 \rangle$ levels, despite the substantial hyperfine structure of ${}^{139}$La$^{2+}$.  Population accumulated in other hyperfine levels, through off-resonant scattering and other processes, can be optically pumped to this cycling transition by driving each ${}^2D$ $| F = n \rangle$ to ${}^2F^o$ $| F = n+1 \rangle$ transition, as illustrated in Fig.~\ref{fig:cooling_level_diagram}(c).  The additional required frequencies may be produced by wide-bandwidth fiber electro-optic modulators (standard components at these wavelengths) driven at multiple frequencies to generate sidebands on the laser light for depopulating all hyperfine levels in ${}^2D_{3/2}$ and ${}^2D_{5/2}$.  Therefore, two lasers and two fiber electro-optic modulators can produce all required frequencies for laser cooling.

The aforementioned lifetime of the ${}^2F^o_{5/2}$ and ${}^2F^o_{7/2}$ levels~\cite{safronova2014:laiii} results in a Doppler-cooling temperature limit of less than 1~$\mu$K, which might make La$^{2+}$ ions attractive for setups involving sympathetic cooling.  On the other hand, these relatively long lifetimes also limit the photon scattering rate from La$^{2+}$ ions, which is important for fluorescence detection of trapped ions and state detection of a potential La$^{2+}$ hyperfine qubit.

%%%%%%%%%%%%%%%%%%%%%%%%%%%%%%%%%%%%%%%%%%
\section{Hyperfine qubit}
	\label{sec:qubit}
Qubits stored in the hyperfine levels of a trapped ion can exhibit simple state preparation, long coherence times, and high-fidelity state detection (for a review of different qubit realizations in trapped atomic ions, see e.g.~\cite{ozeri2011:trapped_ion_tool_box, bruzewicz2019:trapped_ion_qc_review}).  We consider a hyperfine qubit in ${}^{139}$La$^{2+}$ stored in states in the ${}^2D_{5/2}$ $| F = 5 \rangle$ and $| F = 6 \rangle$ hyperfine manifolds.

Initialization of a qubit into a defined pure state is the first step in most quantum information protocols.  Doubly-ionized lanthanum could be prepared in the ${}^2D_{5/2}$ $| F = 6, m_F = +6 \rangle$ or $| F = 6, m_F = -6 \rangle$ state by optical pumping with $\sigma^+$- or $\sigma^-$-polarized light, respectively.  Optical pumping with $\sigma^-$-polarized light to $| F = 6, m_F = -6 \rangle$ may be preferred to red-detune the incident laser light from all ${}^2D_{5/2}$ $| F = 5, 6 \rangle$ $\leftrightarrow$ ${}^2F^o_{7/2}$ $| F = 6, 7 \rangle$ transitions while optimizing the scattering rate on the cycling transition.  Although state preparation fidelity based on optical pumping in hyperfine qubits with nuclear spin greater than 1/2 is typically limited by the polarization purity of the laser light, it can be improved by combining optical pumping with additional optical or microwave transitions~\cite{benhelm2008:ca43, harty2014:high-fidelity_ion_qubits, gaebler2016:high-fidelity_gate_set_Be_ions}.  A subsequent series of microwave transitions could then move population to any state in the hyperfine manifold, including pairs of states that are first-order insensitive to magnetic field fluctuations.

\begin{figure}
  \includegraphics[width=1.0\columnwidth,keepaspectratio]{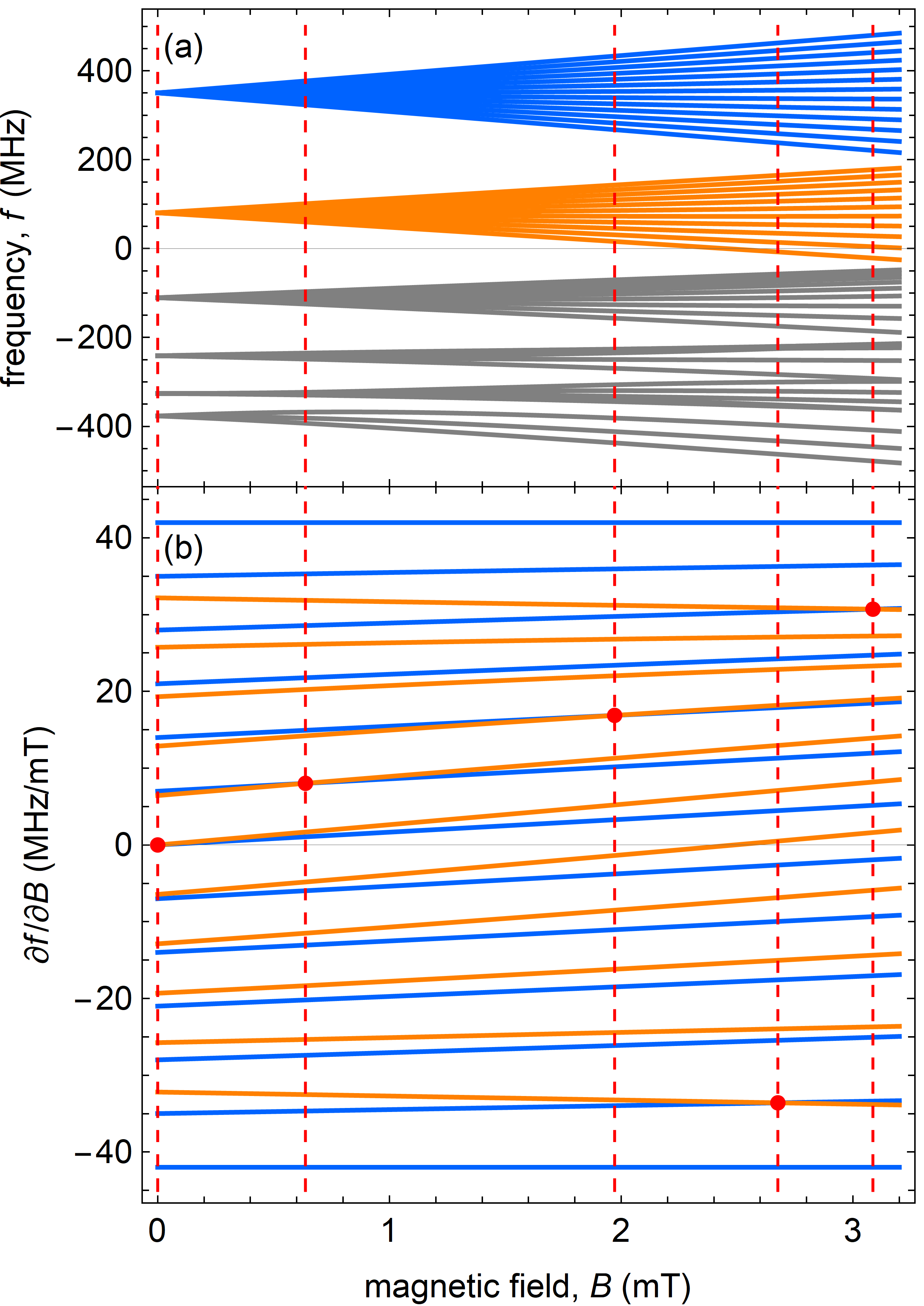}
\caption{Calculating the hyperfine interaction in the presence of an external magnetic field is used to identify possible qubit states with splitting first-order insensitive to the magnetic field.  (a) Frequency shift ($f$) of states in ${}^2D_{5/2}$ as a function of magnetic field ($B$), using the the theoretical hyperfine coefficients from Ref.~\cite{li2021:hyperfine_laiii}.  (b) First-order derivative of the $F = 5, 6$ frequency shifts in (a) with respect to the magnetic field ($\partial f / \partial B$).  Blue lines are states in the $F = 6$ manifold, orange lines are states in the $F = 5$ manifold, red points are where two states have identical derivatives and indicate possible first-order magnetic field insensitve qubits (see Table~\ref{table:qubit_states}), and vertical red dashed lines highlight the strength of the external magnetic field at these points.}
\label{fig:mag_insen_pts}
\end{figure}

Qubits that are first-order insensitive to magnetic fields are important for achieving long coherence times.  We calculated the energy shift of states in the low-lying levels of ${}^{139}$La$^{2+}$ by numerically diagonalizing the Hamiltonian for the hyperfine interaction in the presence of an external magnetic field~\cite{arimondo1977:hyperfine}:
\begin{eqnarray*}
H_{\text{hfs}} & = & A_{\text{hfs}} I \cdot J \\
		& & + B_{\text{hfs}} \frac{6 (I \cdot J)^2 + 3 (I \cdot J) -2 I (I+1) J (J+1)}{2 I (2 I - 1) 2 J (2 J - 1)} \\
		& & + g_J \mu_B B J_z + g_I \mu_B B I_z
\end{eqnarray*}
where $A_{\text{hfs}}$ is the magnetic dipole hyperfine coefficient, $B_{\text{hfs}}$ is the electric quadrupole hyperfine coefficient, $I$ is the nuclear spin and $I_z$ its projection along the magnetic field, $J$ is the total electron spin and $J_z$ its projection along the magnetic field, $\mu_B$ is the Bohr magneton, $B$ is the magnetic field defined to be in the $z$-direction, $g_J$ is the Land\'{e} g-factor, and $g_I$ is the nuclear g-factor using the convention of Ref.~\cite{arimondo1977:hyperfine}.  We repeated the calculation for both the experimentally~\cite{olmschenk2017:laiii_hyperfine} and theoretically~\cite{li2021:hyperfine_laiii} determined values for the hyperfine coefficients, and searched for pairs of states in the ${}^2D_{5/2}$ $| F = 5 \rangle$ and $| F = 6 \rangle$ hyperfine manifolds where the first-order derivative of the frequency difference vanishes at finite magnetic field (see Fig.~\ref{fig:mag_insen_pts}).  Several options for a first-order magnetic field-insensitive qubit are listed in Table~\ref{table:qubit_states}.  

% Table
\begin{table}[b]%The best place to locate the table environment is directly after its first reference in text
	\caption{\label{table:qubit_states}%
	Partial list of options for qubits in the ${}^2D_{5/2}$ $| F = 6 \rangle$ and $| F = 5 \rangle$ hyperfine manifolds with energy splittings that are first-order insensitive to magnetic fields at relatively small fields.  The first column gives the pair of states comprising the qubit, where states are denoted as $| F , m_F \rangle$.  The second column is the value of the magnetic field (in mT) where first-order dependence vanishes.  The third column is the remaining second-order sensitivity to the magnetic field (in Hz/$\mu \text{T}^2$).  In both the second and third columns, the first value is calculated using the experimentally determined hyperfine coefficients from Ref.~\cite{olmschenk2017:laiii_hyperfine} with an estimated uncertainty determined by propagating the experimental uncertainties in the hyperfine coefficients, and the second (parenthetical) value is calculated using the theoretical hyperfine coefficients from Ref.~\cite{li2021:hyperfine_laiii}.
	}
	\begin{ruledtabular}
		\begin{tabular}{lll}
			\textrm{Qubit states}&
			\textrm{Mag. field (mT)}&
			\textrm{Sens. (Hz/$\mu \text{T}^2$)}\\
			\colrule
			$| 5, 0 \rangle \leftrightarrow | 6, 0 \rangle$   & 0. 						& $1.7 \pm 0.8$ (0.49) \\
			$| 5, 1 \rangle \leftrightarrow | 6, 1 \rangle$   & $0.18 \pm 0.07$ (0.64) 	& $1.5 \pm 0.8$ (0.43) \\
			$| 5, 2 \rangle \leftrightarrow | 6, 2 \rangle$   & $0.4  \pm 0.2$ (2.0) 	& $1.2 \pm 0.7$ (0.23) \\
			$| 5, -5 \rangle \leftrightarrow | 6, -5 \rangle$ & $2.0  \pm 0.4$ (2.7) 	& $0.7 \pm 0.2$ (0.53) \\
			$| 5, 5 \rangle \leftrightarrow | 6, 4 \rangle$   & $2.3  \pm 0.4$ (3.1) 	& $0.9 \pm 0.2$ (0.64) \\
		\end{tabular}
	\end{ruledtabular}
\end{table}
%

% Figure

A qubit stored in ${}^2D_{5/2}$ $| F = 5, m_F = 1 \rangle$ and $| F = 6, m_F = 1 \rangle$, calculated to be first-order magnetic field-insensitive at an external field of about 0.18 (0.64) mT using the experimentally (theoretically) determined hyperfine coefficients, is of particular interest.  In this case, the small external magnetic field required is easily obtained in the lab.  Additionally, at this magnetic field the induced shift in transition frequencies from ${}^2D_{5/2}$ $| F = 6 \rangle$ to ${}^2F^o_{7/2}$ $| F = 7 \rangle$ for $\sigma$-polarized light span about 2 (8) MHz across all states, making the manifold addressable by a single laser frequency of reasonable ($\sim 1$ MHz) linewidth.  As noted in Table~\ref{table:qubit_states}, the splitting between this pair of states is calculated to have a second-order magnetic field dependence of about 1.5 Hz/$\mu \text{T}^2$ (0.43 Hz/$\mu \text{T}^2$), comparable to the sensitivity of other qubits with demonstrated long coherence times~\cite{langer2005:long-lived_qubit}.

The qubit state could be read out using standard state-dependent fluorescence~\cite{wineland:nist-jnl, ozeri2011:trapped_ion_tool_box, bruzewicz2019:trapped_ion_qc_review}.  In this case, near-resonant, $\sigma^{-}$-polarized,  laser light could drive the ${}^2D_{5/2}$ $| F = 6 \rangle$ to ${}^2F^o_{7/2}$ $| F = 7 \rangle$ transition, so that population in ${}^2D_{5/2}$ $| F = 6, m_F = 1 \rangle$ qubit state is quickly driven back to the cycling transition, scattering many photons and appearing ``bright" to a detector.  The light used for detection would be detuned from a transition involving the ${}^2D_{5/2}$ $| F = 5, m_F = 1 \rangle$ qubit state by the difference between the ${}^2F^o_{7/2}$ $| F = 6-7 \rangle$ hyperfine splitting and the ${}^2D_{5/2}$ $| F = 5-6 \rangle$ hyperfine splitting, and therefore appear ``dark" to a detector.  Modeling the system similar to Ref.~\cite{acton2006:detection, langer2006:thesis}, we find that errors in state detection due to off-resonant coupling to other states should be small, due to the relatively large detuning with respect to linewidth for both the bright and dark states.  Even when considering the expected linewidth of the detection laser (1 MHz), the detuning to unintended states is approximately $10^3$ linewidths, which is larger than the detuning in several other ions with demonstrated high-fidelity state detection of hyperfine qubits (e.g. ${}^{111}$Cd${}^+$ with off-resonant coupling from the dark state detuned by about $2 \times 10^2$ linewidths~\cite{acton2006:detection}; ${}^{171}$Yb${}^+$ with off-resonant coupling from the bright state detuned by about $10^2$ linewidths~\cite{olmschenk:state-detect, noek2013:state_detect_yb}).  Leakage out of the bright and dark state manifold is further suppressed by small electric dipole transition probabilities between the off-resonantly coupled states, relative to the cycling transition.  Error in bright state detection is also suppressed due to the dependence on polarization errors; either $\pi$- or $\sigma^{+}$-polarized light is required to couple ${}^2D_{5/2}$ $| F = 6, m_F = -6 \rangle$ to off-resonant states instead of driving the cycling transition.  However, decay of the ${}^2D_{5/2}$ level appears as an additional error in bright state detection.  Another important consideration in our model is the error contribution from background counts, given the small photon scattering rate resulting from the relatively long lifetime of the ${}^2F^o_{7/2}$ level.  Nevertheless, given the high detection efficiency and low dark count rate of available superconducting nanowire single-photon detectors (SNSPDs), high state detection fidelities may still be achieved.  Altogether, assuming a simple threshold detection method, modest light collection optics ($NA = 0.28$), a wide range of laser intensities ($I/I_{sat} \lesssim 15$), and detection parameters readily achieved with an SNSPD ($QE \approx 0.8$, dark count rate $\leq 2$ s${}^{-1}$, fiber coupling $\approx 0.2$), our model already predicts a state detection fidelity greater than $99.9 \%$.  The fidelity increases by considering larger light collection~\cite{acton2006:detection, olmschenk:state-detect, maiwald2012:collecting_photons_ion, noek2013:state_detect_yb, chou2017:parabolic, ghadimi2017:diffractive_mirrors, crocker2019:pure_single_photons, araneda2020:panopticon_device, meraner2020:ion_network_node, walker2020:photons_ion_cavity, takahashi2020:ion_cavity_coupling, schupp2021:ion_photon_cavity, kobel2021:ion_photon_cavity}, larger fiber coupling~\cite{stephenson2020:remote_entanglement}, or a smaller laser linewidth (e.g. comparable to the natural linewidth of the transition).  

The potential for first-order magnetic field-insensitive qubit states with high-fidelity state preparation and readout supports ${}^{139}$La$^{2+}$ as an excellent quantum memory candidate, with possible applications in photon-mediated entanglement operations.

%%%%%%%%%%%%%%%%%%%%%%%%%%%%%%%%%%%%%%%%%%
\section{Photonic entanglement}
	\label{sec:entanglement}
Entanglement between an atomic ion and an emitted photon can serve as the foundation for a quantum network, where linking ions over a long-distance is accomplished by the interference and detection of spontaneously emitted photons~\cite{briegel1998:quantum_repeater, simon2003:dist_entangle, kimble2008:qinternet, sangouard2009:q-repeater_ions, duan2010:quantum_networks_ions, awschalom2021:quantum_interconnects}.  Doubly-ionized lanthanum is an excellent candidate for this architecture, as it could be entangled with the polarization or frequency mode of a photon with emitted wavelengths directly amenable to long-distance transmission. 

\begin{figure*}
  \includegraphics[width=2.0\columnwidth,keepaspectratio]{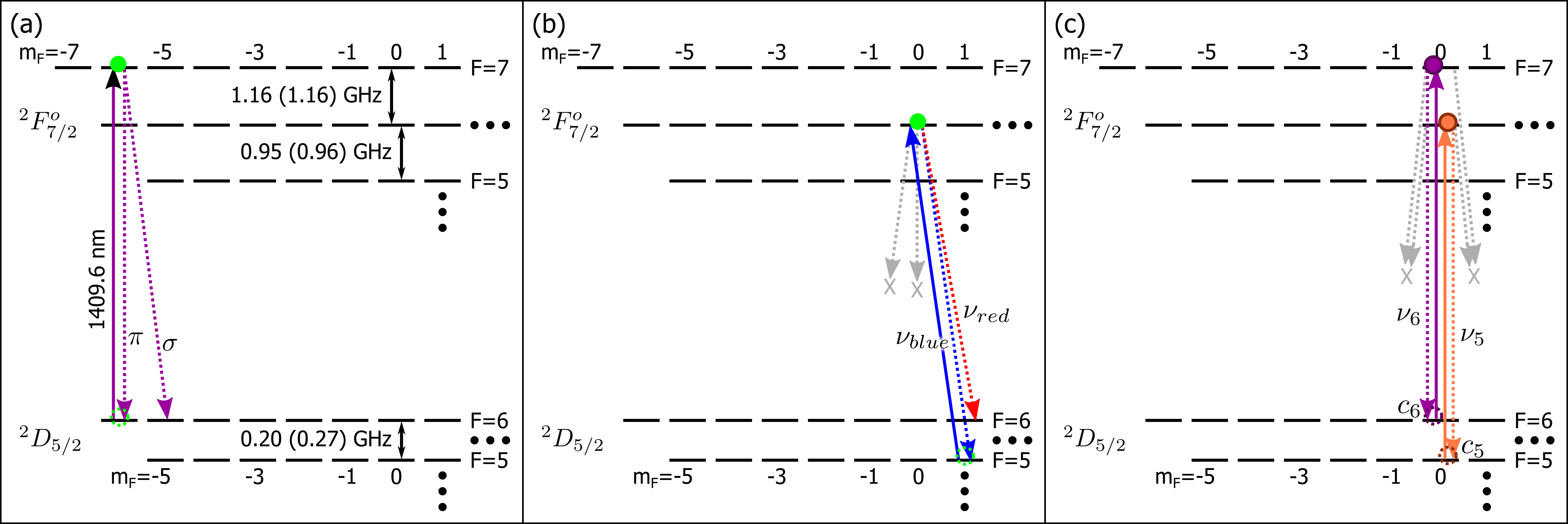}
\caption{Ion-photon entanglement protocols discussed in the text, where dotted circles represent the initially prepared atomic state(s), filled circles represent the state(s) excited by the laser pulse, solid lines represent the transition(s) driven by the laser pulse, dotted lines represent decay (photon emission) transitions, and the hyperfine splittings are calculated from the experimental (theoretical) hyperfine $A$ and $B$ coefficients given in Ref.~\cite{olmschenk2017:laiii_hyperfine} (Ref.~\cite{li2021:hyperfine_laiii}).  Energy-level diagram is partial and not to scale.  Wavelength given in air~\cite{NIST:ASD_2020}  (a) Polarization-encoded photonic qubit:  The polarization of the photon (here $\pi$ and $\sigma$) is entangled with the final atomic state.  Collecting photons in a direction perpendicular to a quantization axis defined by an external magnetic field, $\pi$ and $\sigma$ polarizations are linearly polarized and orthogonal.  (b) Frequency-encoded photonic qubit:  The (resolved) frequency of the emitted photon (here $\nu_{red}$ and $\nu_{blue}$) is entangled with the final atomic state.  Collecting photons along a quantization axis defined by an external magnetic field allows filtering-out $\pi$-polarized light (vertical, gray, dotted line with X) based on the dipole radiation pattern, and $\sigma^+$-polarized light (diagonal, gray, dotted line with X) using a quarter-wave plate and polarizer.  (c) Amplitude-preserving, frequency-encoded photonic qubit:  The (resolved) frequency of the emitted photon (here $\nu_{5}$ and $\nu_{6}$) is entangled with the final atomic state.  Collecting photons in a direction perpendicular to a quantization axis defined by an external magnetic field, $\pi$ and $\sigma$ polarizations are linearly polarized and orthogonal, allowing $\sigma$-polarized light (diagonal, gray, dotted lines with X) to be filtered-out.  Conditioned on this filtering, and due to selection rules, the initially prepared atomic amplitudes $c_5$ and $c_6$ may be preserved.}
\label{fig:ion-photon}       % Give a unique label
\end{figure*}

A possible protocol for entangling ${}^{139}$La$^{2+}$ with the polarization of an emitted photon is illustrated in Fig.~\ref{fig:ion-photon}(a).  The ion would be initially prepared in the ${}^2D_{5/2}$ $| F = 6, m_F = -6 \rangle$ state by optical pumping with $\sigma^-$-polarized light.  A 10 ns, $\pi$-polarized laser pulse then could excite the atom to the ${}^2F^o_{7/2}$ $| F = 7, m_F = -6 \rangle$ state.  Notably, an excitation pulse of 10 ns would be much shorter than the lifetime of the excited state (inhibiting multiple excitation-emission processes), yet result in a pulse linewidth significantly smaller than the excited state hyperfine splitting (inhibiting off-resonant coupling to other states).  We estimate the optimal pulse duration is about 10 ns (though this value will depend on the time-bandwidth product of the pulse), with estimated errors due to double-excitation and off-resonant scattering each less than $3 \times 10^{-3}$.  Following excitation, the subsequent atomic decay and photon emission ideally produces the ion-photon entangled state $|\Psi \rangle = \sqrt{\frac{1}{7}} | 6, -6 \rangle | \pi \rangle + \sqrt{\frac{6}{7}} |6, -5 \rangle | \sigma \rangle$, where the atomic state is denoted by the final $F$ and $m_F$ values in the ${}^2D_{5/2}$ level, the photonic state is denoted by the polarization $\pi$ or $\sigma$, and the coefficients arise from the dipole transition matrix element for each decay channel.  Collecting emitted photons in a direction perpendicular to a quantization axis defined by an external magnetic field, where $\pi$ and $\sigma$ polarizations are linearly polarized and orthogonal, and the intensity of radiation of the $\pi$-decay is twice as strong as the $\sigma$-decay, would result in an ion-photon entangled state $|\Psi \rangle = \frac{1}{2} | 6, -6 \rangle | V \rangle + \frac{\sqrt{3}}{2} | 6, -5 \rangle | H \rangle$, where we now denote the two (linear, orthogonal) polarizations as $V$ and $H$~\cite{blinov2004:ion-photon, moehring2007:review}.  This protocol has the advantage of simple state preparation and straight-forward manipulation of the photonic state using waveplates and polarizers, although it does not directly connect to a magnetic-field insensitive pair of states for the ion.

% Figure

One option for entangling a first-order magnetic field-insensitive atomic qubit with the frequency of an emitted photon is shown in Fig.~\ref{fig:ion-photon}(b).  In this case, following optical pumping to ${}^2D_{5/2}$ $| F = 6, m_F = -6 \rangle$, a series of microwave transitions could move the population to ${}^2D_{5/2}$ $| F = 5, m_F = 1 \rangle$.  A 10 ns, $\sigma^-$-polarized laser pulse then could excite the atom to the ${}^2F^o_{7/2}$ $| F = 6, m_F = 0 \rangle$ state, with estimated errors due to double-excitation and off-resonant scattering nearly the same as the polarization-encoded protocol.  Collecting photons in a direction along a quantization axis defined by an external magnetic field, the dipole radiation pattern of $\pi$-polarized light does not couple into a single-mode optical fiber, and $\sigma^+$- and $\sigma^-$-transitions produce light with orthogonal circular polarizations that may be converted to orthogonal linear polarizations using quarter-wave plates~\cite{moehring2007:review, luo2009:protocols_q_network, kim2011:collect_single_photons_ion, crocker2019:pure_single_photons}.  The atomic decay and photon emission, combined with appropriate filtering to select only the $\sigma^-$-polarized photons, then ideally produces the ion-photon entangled state $| \Psi \rangle = \sqrt{\frac{25}{36}} | 5, 1 \rangle | \nu_{blue} \rangle + \sqrt{\frac{11}{36}} | 6, 1 \rangle | \nu_{red} \rangle$, where the atomic state is again denoted by the final $F$ and $m_F$ values in the ${}^2D_{5/2}$ level and the photonic state is denoted by the frequency $\nu_{blue}$ or $\nu_{red}$, which approximately differ by the hyperfine splitting of $F = 5$ and $F = 6$ in ${}^2D_{5/2}$ ($\nu_{blue} - \nu_{red} = 0.20$ $(0.27)$ GHz) and are well-resolved due to the narrow linewidth of the transition ($\gamma = 2 \pi \times 36$ kHz).  These resulting atomic states can be first-order magnetic field-insensitive (see Table~\ref{table:qubit_states}), and the frequency-encoded photonic states should be resilient against dispersion and birefringence~\cite{duan2006:freq-qubit, moehring2007:review}.

An alternative frequency-encoded photonic qubit protocol is presented in Fig.\ref{fig:ion-photon}(c), where the ion is prepared in a superposition of ${}^2D_{5/2}$ $| F = 5, m_F = 0 \rangle$ and $| F = 6, m_F = 0 \rangle$, then excited to ${}^2F^o_{7/2}$ using $\pi$-polarized light.  Driving the ion with two frequencies, and noting the elimination of nearest-state off-resonant coupling due to selection rules, a few-nanosecond laser pulse could simultaneouly drive the ${}^2D_{5/2}$ $| F = 5, m_F = 0 \rangle$ to ${}^2F^o_{7/2}$ $| F = 6, m_F = 0 \rangle$ and ${}^2D_{5/2}$ $| F = 6, m_F = 0 \rangle$ to ${}^2F^o_{7/2}$ $| F = 7, m_F = 0 \rangle$ transitions with estimated errors due to double-excitation and off-resonant scattering each less than $1 \times 10^{-3}$.  Collecting photons in a direction perpendicular to a quantization axis defined by an external magnetic field allows for polarizaton filtering to select only photons from $\pi$-transitions.  Then due to selection rules, the resulting ion-photon entangled state is ideally $| \Psi \rangle = a_5 c_5 | 5, 0 \rangle | \nu_5 \rangle + a_6 c_6 | 6, 0 \rangle | \nu_6 \rangle$, where the atomic state is again denoted by the final $F$ and $m_F$ values in the ${}^2D_{5/2}$ level, the photonic state is denoted by the resolved frequency $\nu_5$ or $\nu_6$ such that $\nu_6 - \nu_5 = 0.96$ (0.89) GHz (the difference between the ${}^2F^o_{7/2}$ $F = 6 \leftrightarrow 7$ and ${}^2D_{5/2}$ $F = 5 \leftrightarrow 6$ hyperfine splittings), the initial amplitudes of ${}^2D_{5/2}$ $| F = 5, m_F = 0 \rangle$ and $| F = 6, m_F = 0 \rangle$ are denoted by $c_5$ and $c_6$, and the effect of the excitation laser pulse and dipole transition matrix element are incorporated into $a_5$ and $a_6$.  Notably, it should be possible to tailor the excitation laser pulse properties such that $a_5$ and $a_6$ can be assigned arbitrary relative values, including $a_5 = a_6 = 1$.  In that case, the ion-photon entangled state becomes $| \Psi \rangle = c_5 | 5, 0 \rangle | \nu_5 \rangle + c_6 | 6, 0 \rangle | \nu_6 \rangle$, and may facilitate a photon-mediated quantum gate between remote atoms~\cite{duan2006:freq-qubit, olmschenk:teleportation, maunz2009:heralded_gate}.

Additional ion-photon entanglement protocols can be devised to connect to other potential atomic qubits states in Table~\ref{table:qubit_states} (and more beyond these).  Moreover, other photonic qubits can be considered, such as using two successive excitations of the cycling transition to produce a time-bin encoded photonic qubit~\cite{brendel999:time-bin_qubit, barrett2005:time-bin_qubit, luo2009:protocols_q_network, bernien2013:remote_nv_entanglement}.  In all cases, photons at this wavelength are expected to experience $\leq 0.32$ dB/km attenuation in standard optical fiber~\cite{fiber_attenuation_note, smf28ultra_datasheet}.

The rate of ion-photon entanglement is an also important consideration for large-scale quantum networks.  Although the per atom repetition rate for producing ion-photon entanglement is bounded by the relatively long lifetime of the ${}^2F^o_{7/2}$ excited state in ${}^{139}$La$^{2+}$ ($\tau = 4.4$ $\mu$s~\cite{safronova2014:laiii}; $\gamma = 2 \pi \times 36$ kHz), for distances greater than about 5 km between network nodes, the limitation to the per atom repetition rate will be the propagation delay for the photon and (classical) detection signal~\cite{krutyanskiy2019:light-matter_50km, rate_limit_note}.  As with any trapped ion system facing this distance-limited rate, the probability of successfully registering entanglement over long distances can be improved by incorporating high numerical aperture optics~\cite{noek2013:state_detect_yb, chou2017:parabolic, ghadimi2017:diffractive_mirrors, crocker2019:pure_single_photons, araneda2020:panopticon_device} or an optical cavity~\cite{meraner2020:ion_network_node, walker2020:photons_ion_cavity, takahashi2020:ion_cavity_coupling, schupp2021:ion_photon_cavity, kobel2021:ion_photon_cavity} for efficient photon collection, and the effective rate can be increased by multiplexing a system with multiple communication ions~\cite{monroe2014:modular_qc, brown2016:co-design_tiqc, santra2019:q_repeater_two_species, meraner2020:ion_network_node, dhara2021:q_repeater}.  Therefore, doubly-ionized lanthanum should be able to provide ion-photon entanglement rates over long distances comparable to other ions, and may have a distinct advantage by directly producing the telecom-compatible photons amenable to transmission over these distances.

%%%%%%%%%%%%%%%%%%%%%%%%%%%%%%%%%%%%%%%%%%
\section{Conclusion}
\label{sec:conclusion}
Doubly-ionized lanthanum may allow for efficient laser cooling using infrared diode lasers, a range first-order magnetic-field insensitive qubit states with high-fidelity state detection, and viable protocols for directly entangling trapped ions with telecom-compatible photons for long-distance transmission in optical fiber.  The unique features of doubly-ionized lanthanum may also make it a candidate for other quantum information applications, including as a refrigerant ion~\cite{kielpinski2000:sympathetic_cooling, home2009:complete_methods_qip, lekitsch2017:blueprint_tiqc, raghunandan2020:qsim_sympathetic_cooling} (given the small Doppler-cooling temperature limit and large difference in wavelength compared to other trapped ions) or multi-state qudit candidate~\cite{klimov2003:qutrit_ions, senko2015:int_spin_chain, randall2015:microwave_dressed_qubits_qutrits, leupold2018:q_contextual, low2020:qudit_ions} (given the rich hyperfine structure).   While the results presented here are promising, more precise measurements of the hyperfine structure of doubly-ionized lanthanum, as well as additional investigations into producing and trapping multiply-charged ions, may be needed to realize the potential of doubly-ionized lanthanum for applications in quantum information and quantum networks.

%%%%%%%%%%%%%%%%%%%%%%%%%%%%%%%%%%%%%%%%%%
\begin{acknowledgments}
This material is based upon work supported by the National Science Foundation under Grant No. 1752685.
\end{acknowledgments}

%%%%%%%%%%%%%%%%%%%%%%%%%%%%%%%%%%%%%%%%%%
%


\begin{thebibliography}{92}%
\makeatletter
\providecommand \@ifxundefined [1]{%
 \@ifx{#1\undefined}
}%
\providecommand \@ifnum [1]{%
 \ifnum #1\expandafter \@firstoftwo
 \else \expandafter \@secondoftwo
 \fi
}%
\providecommand \@ifx [1]{%
 \ifx #1\expandafter \@firstoftwo
 \else \expandafter \@secondoftwo
 \fi
}%
\providecommand \natexlab [1]{#1}%
\providecommand \enquote  [1]{``#1''}%
\providecommand \bibnamefont  [1]{#1}%
\providecommand \bibfnamefont [1]{#1}%
\providecommand \citenamefont [1]{#1}%
\providecommand \href@noop [0]{\@secondoftwo}%
\providecommand \href [0]{\begingroup \@sanitize@url \@href}%
\providecommand \@href[1]{\@@startlink{#1}\@@href}%
\providecommand \@@href[1]{\endgroup#1\@@endlink}%
\providecommand \@sanitize@url [0]{\catcode `\\12\catcode `\$12\catcode
  `\&12\catcode `\#12\catcode `\^12\catcode `\_12\catcode `\%12\relax}%
\providecommand \@@startlink[1]{}%
\providecommand \@@endlink[0]{}%
\providecommand \url  [0]{\begingroup\@sanitize@url \@url }%
\providecommand \@url [1]{\endgroup\@href {#1}{\urlprefix }}%
\providecommand \urlprefix  [0]{URL }%
\providecommand \Eprint [0]{\href }%
\providecommand \doibase [0]{https://doi.org/}%
\providecommand \selectlanguage [0]{\@gobble}%
\providecommand \bibinfo  [0]{\@secondoftwo}%
\providecommand \bibfield  [0]{\@secondoftwo}%
\providecommand \translation [1]{[#1]}%
\providecommand \BibitemOpen [0]{}%
\providecommand \bibitemStop [0]{}%
\providecommand \bibitemNoStop [0]{.\EOS\space}%
\providecommand \EOS [0]{\spacefactor3000\relax}%
\providecommand \BibitemShut  [1]{\csname bibitem#1\endcsname}%
\let\auto@bib@innerbib\@empty
%</preamble>
\bibitem [{\citenamefont {Langer}\ \emph {et~al.}(2005)\citenamefont {Langer},
  \citenamefont {Ozeri}, \citenamefont {Jost}, \citenamefont {Chiaverini},
  \citenamefont {DeMarco}, \citenamefont {Ben-Kish}, \citenamefont {Blakestad},
  \citenamefont {Britton}, \citenamefont {Hume}, \citenamefont {Itano},
  \citenamefont {Leibfried}, \citenamefont {Reichle}, \citenamefont
  {Rosenband}, \citenamefont {Schaetz}, \citenamefont {Schmidt},\ and\
  \citenamefont {Wineland}}]{langer2005:long-lived_qubit}%
  \BibitemOpen
  \bibfield  {author} {\bibinfo {author} {\bibfnamefont {C.}~\bibnamefont
  {Langer}}, \bibinfo {author} {\bibfnamefont {R.}~\bibnamefont {Ozeri}},
  \bibinfo {author} {\bibfnamefont {J.~D.}\ \bibnamefont {Jost}}, \bibinfo
  {author} {\bibfnamefont {J.}~\bibnamefont {Chiaverini}}, \bibinfo {author}
  {\bibfnamefont {B.}~\bibnamefont {DeMarco}}, \bibinfo {author} {\bibfnamefont
  {A.}~\bibnamefont {Ben-Kish}}, \bibinfo {author} {\bibfnamefont {R.~B.}\
  \bibnamefont {Blakestad}}, \bibinfo {author} {\bibfnamefont {J.}~\bibnamefont
  {Britton}}, \bibinfo {author} {\bibfnamefont {D.~B.}\ \bibnamefont {Hume}},
  \bibinfo {author} {\bibfnamefont {W.~M.}\ \bibnamefont {Itano}}, \bibinfo
  {author} {\bibfnamefont {D.}~\bibnamefont {Leibfried}}, \bibinfo {author}
  {\bibfnamefont {R.}~\bibnamefont {Reichle}}, \bibinfo {author} {\bibfnamefont
  {T.}~\bibnamefont {Rosenband}}, \bibinfo {author} {\bibfnamefont
  {T.}~\bibnamefont {Schaetz}}, \bibinfo {author} {\bibfnamefont {P.~O.}\
  \bibnamefont {Schmidt}},\ and\ \bibinfo {author} {\bibfnamefont {D.~J.}\
  \bibnamefont {Wineland}},\ }\bibfield  {title} {\bibinfo {title} {Long-lived
  qubit memory using atomic ions},\ }\href
  {https://doi.org/10.1103/PhysRevLett.95.060502} {\bibfield  {journal}
  {\bibinfo  {journal} {Phys. Rev. Lett.}\ }\textbf {\bibinfo {volume} {95}},\
  \bibinfo {pages} {060502} (\bibinfo {year} {2005})}\BibitemShut {NoStop}%
\bibitem [{\citenamefont {Harty}\ \emph {et~al.}(2014)\citenamefont {Harty},
  \citenamefont {Allcock}, \citenamefont {Ballance}, \citenamefont {Guidoni},
  \citenamefont {Janacek}, \citenamefont {Linke}, \citenamefont {Stacey},\ and\
  \citenamefont {Lucas}}]{harty2014:high-fidelity_ion_qubits}%
  \BibitemOpen
  \bibfield  {author} {\bibinfo {author} {\bibfnamefont {T.~P.}\ \bibnamefont
  {Harty}}, \bibinfo {author} {\bibfnamefont {D.~T.~C.}\ \bibnamefont
  {Allcock}}, \bibinfo {author} {\bibfnamefont {C.~J.}\ \bibnamefont
  {Ballance}}, \bibinfo {author} {\bibfnamefont {L.}~\bibnamefont {Guidoni}},
  \bibinfo {author} {\bibfnamefont {H.~A.}\ \bibnamefont {Janacek}}, \bibinfo
  {author} {\bibfnamefont {N.~M.}\ \bibnamefont {Linke}}, \bibinfo {author}
  {\bibfnamefont {D.~N.}\ \bibnamefont {Stacey}},\ and\ \bibinfo {author}
  {\bibfnamefont {D.~M.}\ \bibnamefont {Lucas}},\ }\bibfield  {title} {\bibinfo
  {title} {High-fidelity preparation, gates, memory, and readout of a
  trapped-ion quantum bit},\ }\href
  {https://doi.org/10.1103/PhysRevLett.113.220501} {\bibfield  {journal}
  {\bibinfo  {journal} {Phys. Rev. Lett.}\ }\textbf {\bibinfo {volume} {113}},\
  \bibinfo {pages} {220501} (\bibinfo {year} {2014})},\ \Eprint
  {https://arxiv.org/abs/arXiv:1403.1524} {arXiv:1403.1524} \BibitemShut
  {NoStop}%
\bibitem [{\citenamefont {Wang}\ \emph {et~al.}(2017)\citenamefont {Wang},
  \citenamefont {Um}, \citenamefont {Zhang}, \citenamefont {An}, \citenamefont
  {Lyu}, \citenamefont {Zhang}, \citenamefont {Duan}, \citenamefont {Yum},\
  and\ \citenamefont {Kim}}]{wang2017:qubit_coherence}%
  \BibitemOpen
  \bibfield  {author} {\bibinfo {author} {\bibfnamefont {Y.}~\bibnamefont
  {Wang}}, \bibinfo {author} {\bibfnamefont {M.}~\bibnamefont {Um}}, \bibinfo
  {author} {\bibfnamefont {J.}~\bibnamefont {Zhang}}, \bibinfo {author}
  {\bibfnamefont {S.}~\bibnamefont {An}}, \bibinfo {author} {\bibfnamefont
  {M.}~\bibnamefont {Lyu}}, \bibinfo {author} {\bibfnamefont {J.-N.}\
  \bibnamefont {Zhang}}, \bibinfo {author} {\bibfnamefont {L.-M.}\ \bibnamefont
  {Duan}}, \bibinfo {author} {\bibfnamefont {D.}~\bibnamefont {Yum}},\ and\
  \bibinfo {author} {\bibfnamefont {K.}~\bibnamefont {Kim}},\ }\bibfield
  {title} {\bibinfo {title} {Single-qubit quantum memory exceeding ten-minute
  coherence time},\ }\href {https://doi.org/10.1038/s41566-017-0007-1}
  {\bibfield  {journal} {\bibinfo  {journal} {Nature Photonics}\ }\textbf
  {\bibinfo {volume} {11}},\ \bibinfo {pages} {646} (\bibinfo {year} {2017})},\
  \Eprint {https://arxiv.org/abs/arXiv:1701.04195} {arXiv:1701.04195}
  \BibitemShut {NoStop}%
\bibitem [{\citenamefont {Wang}\ \emph {et~al.}(2021)\citenamefont {Wang},
  \citenamefont {Luan}, \citenamefont {Qiao}, \citenamefont {Um}, \citenamefont
  {Zhang}, \citenamefont {Wang}, \citenamefont {Yuan}, \citenamefont {Gu},
  \citenamefont {Zhang},\ and\ \citenamefont {Kim}}]{wang2021:qubit_coherence}%
  \BibitemOpen
  \bibfield  {author} {\bibinfo {author} {\bibfnamefont {P.}~\bibnamefont
  {Wang}}, \bibinfo {author} {\bibfnamefont {C.-Y.}\ \bibnamefont {Luan}},
  \bibinfo {author} {\bibfnamefont {M.}~\bibnamefont {Qiao}}, \bibinfo {author}
  {\bibfnamefont {M.}~\bibnamefont {Um}}, \bibinfo {author} {\bibfnamefont
  {J.}~\bibnamefont {Zhang}}, \bibinfo {author} {\bibfnamefont
  {Y.}~\bibnamefont {Wang}}, \bibinfo {author} {\bibfnamefont {X.}~\bibnamefont
  {Yuan}}, \bibinfo {author} {\bibfnamefont {M.}~\bibnamefont {Gu}}, \bibinfo
  {author} {\bibfnamefont {J.}~\bibnamefont {Zhang}},\ and\ \bibinfo {author}
  {\bibfnamefont {K.}~\bibnamefont {Kim}},\ }\bibfield  {title} {\bibinfo
  {title} {Single ion qubit with estimated coherence time exceeding one hour},\
  }\href {https://doi.org/10.1038/s41467-020-20330-w} {\bibfield  {journal}
  {\bibinfo  {journal} {Nat. Commun.}\ }\textbf {\bibinfo {volume} {12}},\
  \bibinfo {pages} {233} (\bibinfo {year} {2021})},\ \Eprint
  {https://arxiv.org/abs/arXiv:2008.00251} {arXiv:2008.00251} \BibitemShut
  {NoStop}%
\bibitem [{\citenamefont {Gaebler}\ \emph {et~al.}(2016)\citenamefont
  {Gaebler}, \citenamefont {Tan}, \citenamefont {Lin}, \citenamefont {Wan},
  \citenamefont {Bowler}, \citenamefont {Keith}, \citenamefont {Glancy},
  \citenamefont {Coakley}, \citenamefont {Knill}, \citenamefont {Leibfried},\
  and\ \citenamefont {Wineland}}]{gaebler2016:high-fidelity_gate_set_Be_ions}%
  \BibitemOpen
  \bibfield  {author} {\bibinfo {author} {\bibfnamefont {J.~P.}\ \bibnamefont
  {Gaebler}}, \bibinfo {author} {\bibfnamefont {T.~R.}\ \bibnamefont {Tan}},
  \bibinfo {author} {\bibfnamefont {Y.}~\bibnamefont {Lin}}, \bibinfo {author}
  {\bibfnamefont {Y.}~\bibnamefont {Wan}}, \bibinfo {author} {\bibfnamefont
  {R.}~\bibnamefont {Bowler}}, \bibinfo {author} {\bibfnamefont {A.~C.}\
  \bibnamefont {Keith}}, \bibinfo {author} {\bibfnamefont {S.}~\bibnamefont
  {Glancy}}, \bibinfo {author} {\bibfnamefont {K.}~\bibnamefont {Coakley}},
  \bibinfo {author} {\bibfnamefont {E.}~\bibnamefont {Knill}}, \bibinfo
  {author} {\bibfnamefont {D.}~\bibnamefont {Leibfried}},\ and\ \bibinfo
  {author} {\bibfnamefont {D.~J.}\ \bibnamefont {Wineland}},\ }\bibfield
  {title} {\bibinfo {title} {High-fidelity universal gate set for
  ${}^9${Be}${}^+$ ion qubits},\ }\href
  {https://doi.org/10.1103/PhysRevLett.117.060505} {\bibfield  {journal}
  {\bibinfo  {journal} {Phys. Rev. Lett.}\ }\textbf {\bibinfo {volume} {117}},\
  \bibinfo {pages} {060505} (\bibinfo {year} {2016})},\ \Eprint
  {https://arxiv.org/abs/arXiv:1604.00032} {arXiv:1604.00032} \BibitemShut
  {NoStop}%
\bibitem [{\citenamefont {Christensen}\ \emph {et~al.}(2020)\citenamefont
  {Christensen}, \citenamefont {Hucul}, \citenamefont {Campbell},\ and\
  \citenamefont {Hudson}}]{christensen2020:high-fidelity_qubit}%
  \BibitemOpen
  \bibfield  {author} {\bibinfo {author} {\bibfnamefont {J.~E.}\ \bibnamefont
  {Christensen}}, \bibinfo {author} {\bibfnamefont {D.}~\bibnamefont {Hucul}},
  \bibinfo {author} {\bibfnamefont {W.~C.}\ \bibnamefont {Campbell}},\ and\
  \bibinfo {author} {\bibfnamefont {E.~R.}\ \bibnamefont {Hudson}},\ }\bibfield
   {title} {\bibinfo {title} {High-fidelity manipulation of a qubit enabled by
  a manufactured nucleus},\ }\href {https://doi.org/10.1038/s41534-020-0265-5}
  {\bibfield  {journal} {\bibinfo  {journal} {npj Quantum Information}\
  }\textbf {\bibinfo {volume} {6}},\ \bibinfo {pages} {35} (\bibinfo {year}
  {2020})},\ \Eprint {https://arxiv.org/abs/arXiv:1907.13331}
  {arXiv:1907.13331} \BibitemShut {NoStop}%
\bibitem [{\citenamefont {Zhukas}\ \emph {et~al.}(2021)\citenamefont {Zhukas},
  \citenamefont {Svihra}, \citenamefont {Nomerotski},\ and\ \citenamefont
  {Blinov}}]{zhukas2021:detection_qubit_register}%
  \BibitemOpen
  \bibfield  {author} {\bibinfo {author} {\bibfnamefont {L.~A.}\ \bibnamefont
  {Zhukas}}, \bibinfo {author} {\bibfnamefont {P.}~\bibnamefont {Svihra}},
  \bibinfo {author} {\bibfnamefont {A.}~\bibnamefont {Nomerotski}},\ and\
  \bibinfo {author} {\bibfnamefont {B.~B.}\ \bibnamefont {Blinov}},\ }\bibfield
   {title} {\bibinfo {title} {High-fidelity simultaneous detection of a
  trapped-ion qubit register},\ }\href
  {https://doi.org/10.1103/PhysRevA.103.062614} {\bibfield  {journal} {\bibinfo
   {journal} {Phys. Rev. A}\ }\textbf {\bibinfo {volume} {103}},\ \bibinfo
  {pages} {062614} (\bibinfo {year} {2021})},\ \Eprint
  {https://arxiv.org/abs/arXiv:2006.12801} {arXiv:2006.12801} \BibitemShut
  {NoStop}%
\bibitem [{\citenamefont {Blatt}\ and\ \citenamefont
  {Wineland}(2008)}]{blatt2008:ions_entangled}%
  \BibitemOpen
  \bibfield  {author} {\bibinfo {author} {\bibfnamefont {R.}~\bibnamefont
  {Blatt}}\ and\ \bibinfo {author} {\bibfnamefont {D.~J.}\ \bibnamefont
  {Wineland}},\ }\bibfield  {title} {\bibinfo {title} {Entangled states of
  trapped atomic ions},\ }\href {https://doi.org/10.1038/nature07125}
  {\bibfield  {journal} {\bibinfo  {journal} {Nature}\ }\textbf {\bibinfo
  {volume} {453}},\ \bibinfo {pages} {1008} (\bibinfo {year}
  {2008})}\BibitemShut {NoStop}%
\bibitem [{\citenamefont {Bruzewicz}\ \emph {et~al.}(2019)\citenamefont
  {Bruzewicz}, \citenamefont {Chiaverini}, \citenamefont {McConnell},\ and\
  \citenamefont {Sage}}]{bruzewicz2019:trapped_ion_qc_review}%
  \BibitemOpen
  \bibfield  {author} {\bibinfo {author} {\bibfnamefont {C.~D.}\ \bibnamefont
  {Bruzewicz}}, \bibinfo {author} {\bibfnamefont {J.}~\bibnamefont
  {Chiaverini}}, \bibinfo {author} {\bibfnamefont {R.}~\bibnamefont
  {McConnell}},\ and\ \bibinfo {author} {\bibfnamefont {J.~M.}\ \bibnamefont
  {Sage}},\ }\bibfield  {title} {\bibinfo {title} {Trapped-ion quantum
  computing: Progress and challenges},\ }\href
  {https://doi.org/10.1063/1.5088164} {\bibfield  {journal} {\bibinfo
  {journal} {Appl. Phys. Rev.}\ }\textbf {\bibinfo {volume} {6}},\ \bibinfo
  {pages} {021314} (\bibinfo {year} {2019})},\ \Eprint
  {https://arxiv.org/abs/arXiv:1904.04178} {arXiv:1904.04178} \BibitemShut
  {NoStop}%
\bibitem [{\citenamefont {Alexeev}\ \emph {et~al.}(2021)\citenamefont
  {Alexeev}, \citenamefont {Bacon}, \citenamefont {Brown}, \citenamefont
  {Calderbank}, \citenamefont {Carr}, \citenamefont {Chong}, \citenamefont
  {DeMarco}, \citenamefont {Englund}, \citenamefont {Farhi}, \citenamefont
  {Fefferman}, \citenamefont {Gorshkov}, \citenamefont {Houck}, \citenamefont
  {Kim}, \citenamefont {Kimmel}, \citenamefont {Lange}, \citenamefont {Lloyd},
  \citenamefont {Lukin}, \citenamefont {Maslov}, \citenamefont {Maunz},
  \citenamefont {Monroe}, \citenamefont {Preskill}, \citenamefont {Roetteler},
  \citenamefont {Savage},\ and\ \citenamefont
  {Thompson}}]{alexeev2021:q_comp_sci_discovery}%
  \BibitemOpen
  \bibfield  {author} {\bibinfo {author} {\bibfnamefont {Y.}~\bibnamefont
  {Alexeev}}, \bibinfo {author} {\bibfnamefont {D.}~\bibnamefont {Bacon}},
  \bibinfo {author} {\bibfnamefont {K.~R.}\ \bibnamefont {Brown}}, \bibinfo
  {author} {\bibfnamefont {R.}~\bibnamefont {Calderbank}}, \bibinfo {author}
  {\bibfnamefont {L.~D.}\ \bibnamefont {Carr}}, \bibinfo {author}
  {\bibfnamefont {F.~T.}\ \bibnamefont {Chong}}, \bibinfo {author}
  {\bibfnamefont {B.}~\bibnamefont {DeMarco}}, \bibinfo {author} {\bibfnamefont
  {D.}~\bibnamefont {Englund}}, \bibinfo {author} {\bibfnamefont
  {E.}~\bibnamefont {Farhi}}, \bibinfo {author} {\bibfnamefont
  {B.}~\bibnamefont {Fefferman}}, \bibinfo {author} {\bibfnamefont {A.~V.}\
  \bibnamefont {Gorshkov}}, \bibinfo {author} {\bibfnamefont {A.}~\bibnamefont
  {Houck}}, \bibinfo {author} {\bibfnamefont {J.}~\bibnamefont {Kim}}, \bibinfo
  {author} {\bibfnamefont {S.}~\bibnamefont {Kimmel}}, \bibinfo {author}
  {\bibfnamefont {M.}~\bibnamefont {Lange}}, \bibinfo {author} {\bibfnamefont
  {S.}~\bibnamefont {Lloyd}}, \bibinfo {author} {\bibfnamefont {M.~D.}\
  \bibnamefont {Lukin}}, \bibinfo {author} {\bibfnamefont {D.}~\bibnamefont
  {Maslov}}, \bibinfo {author} {\bibfnamefont {P.}~\bibnamefont {Maunz}},
  \bibinfo {author} {\bibfnamefont {C.}~\bibnamefont {Monroe}}, \bibinfo
  {author} {\bibfnamefont {J.}~\bibnamefont {Preskill}}, \bibinfo {author}
  {\bibfnamefont {M.}~\bibnamefont {Roetteler}}, \bibinfo {author}
  {\bibfnamefont {M.~J.}\ \bibnamefont {Savage}},\ and\ \bibinfo {author}
  {\bibfnamefont {J.}~\bibnamefont {Thompson}},\ }\bibfield  {title} {\bibinfo
  {title} {Quantum computer systems for scientific discovery},\ }\href
  {https://doi.org/10.1103/PRXQuantum.2.017001} {\bibfield  {journal} {\bibinfo
   {journal} {PRX Quantum}\ }\textbf {\bibinfo {volume} {2}},\ \bibinfo {pages}
  {017001} (\bibinfo {year} {2021})},\ \Eprint
  {https://arxiv.org/abs/arXiv:1912.07577} {arXiv:1912.07577} \BibitemShut
  {NoStop}%
\bibitem [{\citenamefont {Brown}\ \emph {et~al.}(2021)\citenamefont {Brown},
  \citenamefont {Chiaverini}, \citenamefont {Sage},\ and\ \citenamefont
  {H\"{a}ffner}}]{brown2021:materials_challenges_qc}%
  \BibitemOpen
  \bibfield  {author} {\bibinfo {author} {\bibfnamefont {K.~R.}\ \bibnamefont
  {Brown}}, \bibinfo {author} {\bibfnamefont {J.}~\bibnamefont {Chiaverini}},
  \bibinfo {author} {\bibfnamefont {J.~M.}\ \bibnamefont {Sage}},\ and\
  \bibinfo {author} {\bibfnamefont {H.}~\bibnamefont {H\"{a}ffner}},\
  }\bibfield  {title} {\bibinfo {title} {Materials challenges for trapped-ion
  quantum computers},\ }\href {https://doi.org/10.1038/s41578-021-00292-1}
  {\bibfield  {journal} {\bibinfo  {journal} {Nat. Rev. Mater.}\ } (\bibinfo
  {year} {2021})},\ \Eprint {https://arxiv.org/abs/arXiv:2009.00568}
  {arXiv:2009.00568} \BibitemShut {NoStop}%
\bibitem [{\citenamefont {{de Leon}}\ \emph {et~al.}(2021)\citenamefont {{de
  Leon}}, \citenamefont {Itoh}, \citenamefont {Kim}, \citenamefont {Mehta},
  \citenamefont {Northup}, \citenamefont {Paik}, \citenamefont {Palmer},
  \citenamefont {Samarth}, \citenamefont {Sangtawesin},\ and\ \citenamefont
  {Steuerman}}]{deleon2021:materials_challenges_qc}%
  \BibitemOpen
  \bibfield  {author} {\bibinfo {author} {\bibfnamefont {N.~P.}\ \bibnamefont
  {{de Leon}}}, \bibinfo {author} {\bibfnamefont {K.~M.}\ \bibnamefont {Itoh}},
  \bibinfo {author} {\bibfnamefont {D.}~\bibnamefont {Kim}}, \bibinfo {author}
  {\bibfnamefont {K.~K.}\ \bibnamefont {Mehta}}, \bibinfo {author}
  {\bibfnamefont {T.~E.}\ \bibnamefont {Northup}}, \bibinfo {author}
  {\bibfnamefont {H.}~\bibnamefont {Paik}}, \bibinfo {author} {\bibfnamefont
  {B.~S.}\ \bibnamefont {Palmer}}, \bibinfo {author} {\bibfnamefont
  {N.}~\bibnamefont {Samarth}}, \bibinfo {author} {\bibfnamefont
  {S.}~\bibnamefont {Sangtawesin}},\ and\ \bibinfo {author} {\bibfnamefont
  {D.~W.}\ \bibnamefont {Steuerman}},\ }\bibfield  {title} {\bibinfo {title}
  {Materials challenges and opportunities for quantum computing hardware},\
  }\href {https://doi.org/10.1126/science.abb2823} {\bibfield  {journal}
  {\bibinfo  {journal} {Science}\ }\textbf {\bibinfo {volume} {372}},\ \bibinfo
  {pages} {eabb2823} (\bibinfo {year} {2021})}\BibitemShut {NoStop}%
\bibitem [{\citenamefont {Blatt}\ and\ \citenamefont
  {Roos}(2012)}]{blatt2012:quantum_simulation}%
  \BibitemOpen
  \bibfield  {author} {\bibinfo {author} {\bibfnamefont {R.}~\bibnamefont
  {Blatt}}\ and\ \bibinfo {author} {\bibfnamefont {C.~F.}\ \bibnamefont
  {Roos}},\ }\bibfield  {title} {\bibinfo {title} {Quantum simulations with
  trapped ions},\ }\href {https://doi.org/10.1038/nphys2252} {\bibfield
  {journal} {\bibinfo  {journal} {Nature Physics}\ }\textbf {\bibinfo {volume}
  {8}},\ \bibinfo {pages} {277} (\bibinfo {year} {2012})}\BibitemShut {NoStop}%
\bibitem [{\citenamefont {Monroe}\ \emph {et~al.}(2021)\citenamefont {Monroe},
  \citenamefont {Campbell}, \citenamefont {Duan}, \citenamefont {Gong},
  \citenamefont {Gorshkov}, \citenamefont {Hess}, \citenamefont {Islam},
  \citenamefont {Kim}, \citenamefont {Linke}, \citenamefont {Pagano},
  \citenamefont {Richerme}, \citenamefont {Senko},\ and\ \citenamefont
  {Yao}}]{monroe2021:programmable_quantum_simulations}%
  \BibitemOpen
  \bibfield  {author} {\bibinfo {author} {\bibfnamefont {C.}~\bibnamefont
  {Monroe}}, \bibinfo {author} {\bibfnamefont {W.~C.}\ \bibnamefont
  {Campbell}}, \bibinfo {author} {\bibfnamefont {L.-M.}\ \bibnamefont {Duan}},
  \bibinfo {author} {\bibfnamefont {Z.-X.}\ \bibnamefont {Gong}}, \bibinfo
  {author} {\bibfnamefont {A.~V.}\ \bibnamefont {Gorshkov}}, \bibinfo {author}
  {\bibfnamefont {P.~W.}\ \bibnamefont {Hess}}, \bibinfo {author}
  {\bibfnamefont {R.}~\bibnamefont {Islam}}, \bibinfo {author} {\bibfnamefont
  {K.}~\bibnamefont {Kim}}, \bibinfo {author} {\bibfnamefont {N.~M.}\
  \bibnamefont {Linke}}, \bibinfo {author} {\bibfnamefont {G.}~\bibnamefont
  {Pagano}}, \bibinfo {author} {\bibfnamefont {P.}~\bibnamefont {Richerme}},
  \bibinfo {author} {\bibfnamefont {C.}~\bibnamefont {Senko}},\ and\ \bibinfo
  {author} {\bibfnamefont {N.~Y.}\ \bibnamefont {Yao}},\ }\bibfield  {title}
  {\bibinfo {title} {Programmable quantum simulations of spin systems with
  trapped ions},\ }\href {https://doi.org/10.1103/RevModPhys.93.025001}
  {\bibfield  {journal} {\bibinfo  {journal} {Rev. Mod. Phys.}\ }\textbf
  {\bibinfo {volume} {93}},\ \bibinfo {pages} {025001} (\bibinfo {year}
  {2021})},\ \Eprint {https://arxiv.org/abs/arXiv:1912.07845}
  {arXiv:1912.07845} \BibitemShut {NoStop}%
\bibitem [{\citenamefont {Altman}\ \emph {et~al.}(2021)\citenamefont {Altman},
  \citenamefont {Brown}, \citenamefont {Carleo}, \citenamefont {Carr},
  \citenamefont {Demler}, \citenamefont {Chin}, \citenamefont {DeMarco},
  \citenamefont {Economou}, \citenamefont {Eriksson}, \citenamefont {Fu},
  \citenamefont {Greiner}, \citenamefont {Hazzard}, \citenamefont {Hulet},
  \citenamefont {Koll\'ar}, \citenamefont {Lev}, \citenamefont {Lukin},
  \citenamefont {Ma}, \citenamefont {Mi}, \citenamefont {Misra}, \citenamefont
  {Monroe}, \citenamefont {Murch}, \citenamefont {Nazario}, \citenamefont {Ni},
  \citenamefont {Potter}, \citenamefont {Roushan}, \citenamefont {Saffman},
  \citenamefont {Schleier-Smith}, \citenamefont {Siddiqi}, \citenamefont
  {Simmonds}, \citenamefont {Singh}, \citenamefont {Spielman}, \citenamefont
  {Temme}, \citenamefont {Weiss}, \citenamefont {Vu\ifmmode \check{c}\else
  \v{c}\fi{}kovi\ifmmode~\acute{c}\else \'{c}\fi{}}, \citenamefont
  {Vuleti\ifmmode~\acute{c}\else \'{c}\fi{}}, \citenamefont {Ye},\ and\
  \citenamefont {Zwierlein}}]{altman2021:q_sim_roadmap}%
  \BibitemOpen
  \bibfield  {author} {\bibinfo {author} {\bibfnamefont {E.}~\bibnamefont
  {Altman}}, \bibinfo {author} {\bibfnamefont {K.~R.}\ \bibnamefont {Brown}},
  \bibinfo {author} {\bibfnamefont {G.}~\bibnamefont {Carleo}}, \bibinfo
  {author} {\bibfnamefont {L.~D.}\ \bibnamefont {Carr}}, \bibinfo {author}
  {\bibfnamefont {E.}~\bibnamefont {Demler}}, \bibinfo {author} {\bibfnamefont
  {C.}~\bibnamefont {Chin}}, \bibinfo {author} {\bibfnamefont {B.}~\bibnamefont
  {DeMarco}}, \bibinfo {author} {\bibfnamefont {S.~E.}\ \bibnamefont
  {Economou}}, \bibinfo {author} {\bibfnamefont {M.~A.}\ \bibnamefont
  {Eriksson}}, \bibinfo {author} {\bibfnamefont {K.~C.}\ \bibnamefont {Fu}},
  \bibinfo {author} {\bibfnamefont {M.}~\bibnamefont {Greiner}}, \bibinfo
  {author} {\bibfnamefont {K.~R.}\ \bibnamefont {Hazzard}}, \bibinfo {author}
  {\bibfnamefont {R.~G.}\ \bibnamefont {Hulet}}, \bibinfo {author}
  {\bibfnamefont {A.~J.}\ \bibnamefont {Koll\'ar}}, \bibinfo {author}
  {\bibfnamefont {B.~L.}\ \bibnamefont {Lev}}, \bibinfo {author} {\bibfnamefont
  {M.~D.}\ \bibnamefont {Lukin}}, \bibinfo {author} {\bibfnamefont
  {R.}~\bibnamefont {Ma}}, \bibinfo {author} {\bibfnamefont {X.}~\bibnamefont
  {Mi}}, \bibinfo {author} {\bibfnamefont {S.}~\bibnamefont {Misra}}, \bibinfo
  {author} {\bibfnamefont {C.}~\bibnamefont {Monroe}}, \bibinfo {author}
  {\bibfnamefont {K.}~\bibnamefont {Murch}}, \bibinfo {author} {\bibfnamefont
  {Z.}~\bibnamefont {Nazario}}, \bibinfo {author} {\bibfnamefont {K.-K.}\
  \bibnamefont {Ni}}, \bibinfo {author} {\bibfnamefont {A.~C.}\ \bibnamefont
  {Potter}}, \bibinfo {author} {\bibfnamefont {P.}~\bibnamefont {Roushan}},
  \bibinfo {author} {\bibfnamefont {M.}~\bibnamefont {Saffman}}, \bibinfo
  {author} {\bibfnamefont {M.}~\bibnamefont {Schleier-Smith}}, \bibinfo
  {author} {\bibfnamefont {I.}~\bibnamefont {Siddiqi}}, \bibinfo {author}
  {\bibfnamefont {R.}~\bibnamefont {Simmonds}}, \bibinfo {author}
  {\bibfnamefont {M.}~\bibnamefont {Singh}}, \bibinfo {author} {\bibfnamefont
  {I.}~\bibnamefont {Spielman}}, \bibinfo {author} {\bibfnamefont
  {K.}~\bibnamefont {Temme}}, \bibinfo {author} {\bibfnamefont {D.~S.}\
  \bibnamefont {Weiss}}, \bibinfo {author} {\bibfnamefont {J.}~\bibnamefont
  {Vu\ifmmode \check{c}\else \v{c}\fi{}kovi\ifmmode~\acute{c}\else
  \'{c}\fi{}}}, \bibinfo {author} {\bibfnamefont {V.}~\bibnamefont
  {Vuleti\ifmmode~\acute{c}\else \'{c}\fi{}}}, \bibinfo {author} {\bibfnamefont
  {J.}~\bibnamefont {Ye}},\ and\ \bibinfo {author} {\bibfnamefont
  {M.}~\bibnamefont {Zwierlein}},\ }\bibfield  {title} {\bibinfo {title}
  {Quantum simulators: Architectures and opportunities},\ }\href
  {https://doi.org/10.1103/PRXQuantum.2.017003} {\bibfield  {journal} {\bibinfo
   {journal} {PRX Quantum}\ }\textbf {\bibinfo {volume} {2}},\ \bibinfo {pages}
  {017003} (\bibinfo {year} {2021})},\ \Eprint
  {https://arxiv.org/abs/arXiv:1912.06938} {arXiv:1912.06938} \BibitemShut
  {NoStop}%
\bibitem [{\citenamefont {Ludlow}\ \emph {et~al.}(2015)\citenamefont {Ludlow},
  \citenamefont {Boyd}, \citenamefont {Ye}, \citenamefont {Peik},\ and\
  \citenamefont {Schmidt}}]{ludlow2015:optical_atomic_clocks}%
  \BibitemOpen
  \bibfield  {author} {\bibinfo {author} {\bibfnamefont {A.~D.}\ \bibnamefont
  {Ludlow}}, \bibinfo {author} {\bibfnamefont {M.~M.}\ \bibnamefont {Boyd}},
  \bibinfo {author} {\bibfnamefont {J.}~\bibnamefont {Ye}}, \bibinfo {author}
  {\bibfnamefont {E.}~\bibnamefont {Peik}},\ and\ \bibinfo {author}
  {\bibfnamefont {P.~O.}\ \bibnamefont {Schmidt}},\ }\bibfield  {title}
  {\bibinfo {title} {Optical atomic clocks},\ }\href
  {https://doi.org/10.1103/RevModPhys.87.637} {\bibfield  {journal} {\bibinfo
  {journal} {Rev. Mod. Phys.}\ }\textbf {\bibinfo {volume} {87}},\ \bibinfo
  {pages} {637} (\bibinfo {year} {2015})},\ \Eprint
  {https://arxiv.org/abs/arXiv:1407.3493} {arXiv:1407.3493} \BibitemShut
  {NoStop}%
\bibitem [{\citenamefont {Degen}\ \emph {et~al.}(2017)\citenamefont {Degen},
  \citenamefont {Reinhard},\ and\ \citenamefont
  {Cappellaro}}]{degen2017:quantum_sensing}%
  \BibitemOpen
  \bibfield  {author} {\bibinfo {author} {\bibfnamefont {C.~L.}\ \bibnamefont
  {Degen}}, \bibinfo {author} {\bibfnamefont {F.}~\bibnamefont {Reinhard}},\
  and\ \bibinfo {author} {\bibfnamefont {P.}~\bibnamefont {Cappellaro}},\
  }\bibfield  {title} {\bibinfo {title} {Quantum sensing},\ }\href
  {https://doi.org/10.1103/RevModPhys.89.035002} {\bibfield  {journal}
  {\bibinfo  {journal} {Rev. Mod. Phys.}\ }\textbf {\bibinfo {volume} {89}},\
  \bibinfo {pages} {035002} (\bibinfo {year} {2017})},\ \Eprint
  {https://arxiv.org/abs/arXiv:1611.02427} {arXiv:1611.02427} \BibitemShut
  {NoStop}%
\bibitem [{\citenamefont {Briegel}\ \emph {et~al.}(1998)\citenamefont
  {Briegel}, \citenamefont {{D\"{u}r}}, \citenamefont {Cirac},\ and\
  \citenamefont {Zoller}}]{briegel1998:quantum_repeater}%
  \BibitemOpen
  \bibfield  {author} {\bibinfo {author} {\bibfnamefont {H.-J.}\ \bibnamefont
  {Briegel}}, \bibinfo {author} {\bibfnamefont {W.}~\bibnamefont {{D\"{u}r}}},
  \bibinfo {author} {\bibfnamefont {J.~I.}\ \bibnamefont {Cirac}},\ and\
  \bibinfo {author} {\bibfnamefont {P.}~\bibnamefont {Zoller}},\ }\bibfield
  {title} {\bibinfo {title} {Quantum repeaters: The role of imperfect local
  operations in quantum communication},\ }\href
  {https://doi.org/10.1103/PhysRevLett.81.5932} {\bibfield  {journal} {\bibinfo
   {journal} {Phys. Rev. Lett.}\ }\textbf {\bibinfo {volume} {81}},\ \bibinfo
  {pages} {5932} (\bibinfo {year} {1998})},\ \Eprint
  {https://arxiv.org/abs/arXiv:quant-ph/9803056} {arXiv:quant-ph/9803056}
  \BibitemShut {NoStop}%
\bibitem [{\citenamefont {Simon}\ and\ \citenamefont
  {Irvine}(2003)}]{simon2003:dist_entangle}%
  \BibitemOpen
  \bibfield  {author} {\bibinfo {author} {\bibfnamefont {C.}~\bibnamefont
  {Simon}}\ and\ \bibinfo {author} {\bibfnamefont {W.~T.~M.}\ \bibnamefont
  {Irvine}},\ }\bibfield  {title} {\bibinfo {title} {Robust long-distance
  entanglement and a loophole-free bell test with ions and photons},\ }\href
  {https://doi.org/10.1103/PhysRevLett.91.110405} {\bibfield  {journal}
  {\bibinfo  {journal} {Phys. Rev. Lett.}\ }\textbf {\bibinfo {volume} {91}},\
  \bibinfo {pages} {110405} (\bibinfo {year} {2003})},\ \Eprint
  {https://arxiv.org/abs/arXiv:quant-ph/0303023} {arXiv:quant-ph/0303023}
  \BibitemShut {NoStop}%
\bibitem [{\citenamefont {Kimble}(2008)}]{kimble2008:qinternet}%
  \BibitemOpen
  \bibfield  {author} {\bibinfo {author} {\bibfnamefont {H.~J.}\ \bibnamefont
  {Kimble}},\ }\bibfield  {title} {\bibinfo {title} {The quantum internet},\
  }\href {https://doi.org/10.1038/nature07127} {\bibfield  {journal} {\bibinfo
  {journal} {Nature}\ }\textbf {\bibinfo {volume} {453}},\ \bibinfo {pages}
  {1023} (\bibinfo {year} {2008})},\ \Eprint
  {https://arxiv.org/abs/arXiv:0806.4195} {arXiv:0806.4195} \BibitemShut
  {NoStop}%
\bibitem [{\citenamefont {Sangouard}\ \emph {et~al.}(2009)\citenamefont
  {Sangouard}, \citenamefont {Dubessy},\ and\ \citenamefont
  {Simon}}]{sangouard2009:q-repeater_ions}%
  \BibitemOpen
  \bibfield  {author} {\bibinfo {author} {\bibfnamefont {N.}~\bibnamefont
  {Sangouard}}, \bibinfo {author} {\bibfnamefont {R.}~\bibnamefont {Dubessy}},\
  and\ \bibinfo {author} {\bibfnamefont {C.}~\bibnamefont {Simon}},\ }\bibfield
   {title} {\bibinfo {title} {Quantum repeaters based on single trapped ions},\
  }\href {https://doi.org/10.1103/PhysRevA.79.042340} {\bibfield  {journal}
  {\bibinfo  {journal} {Phys. Rev. A}\ }\textbf {\bibinfo {volume} {79}},\
  \bibinfo {pages} {042340} (\bibinfo {year} {2009})},\ \Eprint
  {https://arxiv.org/abs/arXiv:0902.3127} {arXiv:0902.3127} \BibitemShut
  {NoStop}%
\bibitem [{\citenamefont {Duan}\ and\ \citenamefont
  {Monroe}(2010)}]{duan2010:quantum_networks_ions}%
  \BibitemOpen
  \bibfield  {author} {\bibinfo {author} {\bibfnamefont {L.-M.}\ \bibnamefont
  {Duan}}\ and\ \bibinfo {author} {\bibfnamefont {C.}~\bibnamefont {Monroe}},\
  }\bibfield  {title} {\bibinfo {title} {Colloquium: Quantum networks with
  trapped ions},\ }\href {https://doi.org/10.1103/RevModPhys.82.1209}
  {\bibfield  {journal} {\bibinfo  {journal} {Rev. Mod. Phys.}\ }\textbf
  {\bibinfo {volume} {82}},\ \bibinfo {pages} {1209} (\bibinfo {year}
  {2010})}\BibitemShut {NoStop}%
\bibitem [{\citenamefont {Awschalom}\ \emph {et~al.}(2021)\citenamefont
  {Awschalom}, \citenamefont {Berggren}, \citenamefont {Bernien}, \citenamefont
  {Bhave}, \citenamefont {Carr}, \citenamefont {Davids}, \citenamefont
  {Economou}, \citenamefont {Englund}, \citenamefont {Faraon}, \citenamefont
  {Fejer}, \citenamefont {Guha}, \citenamefont {Gustafsson}, \citenamefont
  {Hu}, \citenamefont {Jiang}, \citenamefont {Kim}, \citenamefont {Korzh},
  \citenamefont {Kumar}, \citenamefont {Kwiat}, \citenamefont
  {Lon\ifmmode~\check{c}\else \v{c}\fi{}ar}, \citenamefont {Lukin},
  \citenamefont {Miller}, \citenamefont {Monroe}, \citenamefont {Nam},
  \citenamefont {Narang}, \citenamefont {Orcutt}, \citenamefont {Raymer},
  \citenamefont {Safavi-Naeini}, \citenamefont {Spiropulu}, \citenamefont
  {Srinivasan}, \citenamefont {Sun}, \citenamefont {Vu\ifmmode \check{c}\else
  \v{c}\fi{}kovi\ifmmode~\acute{c}\else \'{c}\fi{}}, \citenamefont {Waks},
  \citenamefont {Walsworth}, \citenamefont {Weiner},\ and\ \citenamefont
  {Zhang}}]{awschalom2021:quantum_interconnects}%
  \BibitemOpen
  \bibfield  {author} {\bibinfo {author} {\bibfnamefont {D.}~\bibnamefont
  {Awschalom}}, \bibinfo {author} {\bibfnamefont {K.~K.}\ \bibnamefont
  {Berggren}}, \bibinfo {author} {\bibfnamefont {H.}~\bibnamefont {Bernien}},
  \bibinfo {author} {\bibfnamefont {S.}~\bibnamefont {Bhave}}, \bibinfo
  {author} {\bibfnamefont {L.~D.}\ \bibnamefont {Carr}}, \bibinfo {author}
  {\bibfnamefont {P.}~\bibnamefont {Davids}}, \bibinfo {author} {\bibfnamefont
  {S.~E.}\ \bibnamefont {Economou}}, \bibinfo {author} {\bibfnamefont
  {D.}~\bibnamefont {Englund}}, \bibinfo {author} {\bibfnamefont
  {A.}~\bibnamefont {Faraon}}, \bibinfo {author} {\bibfnamefont
  {M.}~\bibnamefont {Fejer}}, \bibinfo {author} {\bibfnamefont
  {S.}~\bibnamefont {Guha}}, \bibinfo {author} {\bibfnamefont {M.~V.}\
  \bibnamefont {Gustafsson}}, \bibinfo {author} {\bibfnamefont
  {E.}~\bibnamefont {Hu}}, \bibinfo {author} {\bibfnamefont {L.}~\bibnamefont
  {Jiang}}, \bibinfo {author} {\bibfnamefont {J.}~\bibnamefont {Kim}}, \bibinfo
  {author} {\bibfnamefont {B.}~\bibnamefont {Korzh}}, \bibinfo {author}
  {\bibfnamefont {P.}~\bibnamefont {Kumar}}, \bibinfo {author} {\bibfnamefont
  {P.~G.}\ \bibnamefont {Kwiat}}, \bibinfo {author} {\bibfnamefont
  {M.}~\bibnamefont {Lon\ifmmode~\check{c}\else \v{c}\fi{}ar}}, \bibinfo
  {author} {\bibfnamefont {M.~D.}\ \bibnamefont {Lukin}}, \bibinfo {author}
  {\bibfnamefont {D.~A.~B.}\ \bibnamefont {Miller}}, \bibinfo {author}
  {\bibfnamefont {C.}~\bibnamefont {Monroe}}, \bibinfo {author} {\bibfnamefont
  {S.~W.}\ \bibnamefont {Nam}}, \bibinfo {author} {\bibfnamefont
  {P.}~\bibnamefont {Narang}}, \bibinfo {author} {\bibfnamefont {J.~S.}\
  \bibnamefont {Orcutt}}, \bibinfo {author} {\bibfnamefont {M.~G.}\
  \bibnamefont {Raymer}}, \bibinfo {author} {\bibfnamefont {A.~H.}\
  \bibnamefont {Safavi-Naeini}}, \bibinfo {author} {\bibfnamefont
  {M.}~\bibnamefont {Spiropulu}}, \bibinfo {author} {\bibfnamefont
  {K.}~\bibnamefont {Srinivasan}}, \bibinfo {author} {\bibfnamefont
  {S.}~\bibnamefont {Sun}}, \bibinfo {author} {\bibfnamefont {J.}~\bibnamefont
  {Vu\ifmmode \check{c}\else \v{c}\fi{}kovi\ifmmode~\acute{c}\else
  \'{c}\fi{}}}, \bibinfo {author} {\bibfnamefont {E.}~\bibnamefont {Waks}},
  \bibinfo {author} {\bibfnamefont {R.}~\bibnamefont {Walsworth}}, \bibinfo
  {author} {\bibfnamefont {A.~M.}\ \bibnamefont {Weiner}},\ and\ \bibinfo
  {author} {\bibfnamefont {Z.}~\bibnamefont {Zhang}},\ }\bibfield  {title}
  {\bibinfo {title} {Development of quantum interconnects {(QuICs)} for
  next-generation information technologies},\ }\href
  {https://doi.org/10.1103/PRXQuantum.2.017002} {\bibfield  {journal} {\bibinfo
   {journal} {PRX Quantum}\ }\textbf {\bibinfo {volume} {2}},\ \bibinfo {pages}
  {017002} (\bibinfo {year} {2021})},\ \Eprint
  {https://arxiv.org/abs/arXiv:1912.06642} {arXiv:1912.06642} \BibitemShut
  {NoStop}%
\bibitem [{\citenamefont {Moehring}\ \emph
  {et~al.}(2007{\natexlab{a}})\citenamefont {Moehring}, \citenamefont {Maunz},
  \citenamefont {Olmschenk}, \citenamefont {Younge}, \citenamefont
  {Matsukevich}, \citenamefont {Duan},\ and\ \citenamefont
  {Monroe}}]{moehring:ion-ion}%
  \BibitemOpen
  \bibfield  {author} {\bibinfo {author} {\bibfnamefont {D.~L.}\ \bibnamefont
  {Moehring}}, \bibinfo {author} {\bibfnamefont {P.}~\bibnamefont {Maunz}},
  \bibinfo {author} {\bibfnamefont {S.}~\bibnamefont {Olmschenk}}, \bibinfo
  {author} {\bibfnamefont {K.~C.}\ \bibnamefont {Younge}}, \bibinfo {author}
  {\bibfnamefont {D.~N.}\ \bibnamefont {Matsukevich}}, \bibinfo {author}
  {\bibfnamefont {L.-M.}\ \bibnamefont {Duan}},\ and\ \bibinfo {author}
  {\bibfnamefont {C.}~\bibnamefont {Monroe}},\ }\bibfield  {title} {\bibinfo
  {title} {Entanglement of single-atom quantum bits at a distance},\ }\href
  {https://doi.org/10.1038/nature06118} {\bibfield  {journal} {\bibinfo
  {journal} {Nature}\ }\textbf {\bibinfo {volume} {449}},\ \bibinfo {pages}
  {68} (\bibinfo {year} {2007}{\natexlab{a}})}\BibitemShut {NoStop}%
\bibitem [{\citenamefont {Olmschenk}\ \emph {et~al.}(2009)\citenamefont
  {Olmschenk}, \citenamefont {Matsukevich}, \citenamefont {Maunz},
  \citenamefont {Hayes}, \citenamefont {Duan},\ and\ \citenamefont
  {Monroe}}]{olmschenk:teleportation}%
  \BibitemOpen
  \bibfield  {author} {\bibinfo {author} {\bibfnamefont {S.}~\bibnamefont
  {Olmschenk}}, \bibinfo {author} {\bibfnamefont {D.~N.}\ \bibnamefont
  {Matsukevich}}, \bibinfo {author} {\bibfnamefont {P.}~\bibnamefont {Maunz}},
  \bibinfo {author} {\bibfnamefont {D.}~\bibnamefont {Hayes}}, \bibinfo
  {author} {\bibfnamefont {L.-M.}\ \bibnamefont {Duan}},\ and\ \bibinfo
  {author} {\bibfnamefont {C.}~\bibnamefont {Monroe}},\ }\bibfield  {title}
  {\bibinfo {title} {Quantum teleportation between distant matter qubits},\
  }\href {https://doi.org/10.1126/science.1167209} {\bibfield  {journal}
  {\bibinfo  {journal} {Science}\ }\textbf {\bibinfo {volume} {323}},\ \bibinfo
  {pages} {486} (\bibinfo {year} {2009})},\ \Eprint
  {https://arxiv.org/abs/arXiv:0907.5240} {arXiv:0907.5240} \BibitemShut
  {NoStop}%
\bibitem [{\citenamefont {Hucul}\ \emph {et~al.}(2015)\citenamefont {Hucul},
  \citenamefont {Inlek}, \citenamefont {Vittorini}, \citenamefont {Crocker},
  \citenamefont {Debnath}, \citenamefont {Clark},\ and\ \citenamefont
  {Monroe}}]{hucul2015:modular_entanglement}%
  \BibitemOpen
  \bibfield  {author} {\bibinfo {author} {\bibfnamefont {D.}~\bibnamefont
  {Hucul}}, \bibinfo {author} {\bibfnamefont {I.~V.}\ \bibnamefont {Inlek}},
  \bibinfo {author} {\bibfnamefont {G.}~\bibnamefont {Vittorini}}, \bibinfo
  {author} {\bibfnamefont {C.}~\bibnamefont {Crocker}}, \bibinfo {author}
  {\bibfnamefont {S.}~\bibnamefont {Debnath}}, \bibinfo {author} {\bibfnamefont
  {S.~M.}\ \bibnamefont {Clark}},\ and\ \bibinfo {author} {\bibfnamefont
  {C.}~\bibnamefont {Monroe}},\ }\bibfield  {title} {\bibinfo {title} {Modular
  entanglement of atomic qubits using photons and phonons},\ }\href
  {https://doi.org/10.1038/nphys3150} {\bibfield  {journal} {\bibinfo
  {journal} {Nature Phys.}\ }\textbf {\bibinfo {volume} {11}},\ \bibinfo
  {pages} {37} (\bibinfo {year} {2015})},\ \Eprint
  {https://arxiv.org/abs/arXiv:1403.3696} {arXiv:1403.3696} \BibitemShut
  {NoStop}%
\bibitem [{\citenamefont {Stephenson}\ \emph {et~al.}(2020)\citenamefont
  {Stephenson}, \citenamefont {Nadlinger}, \citenamefont {Nichol},
  \citenamefont {An}, \citenamefont {Drmota}, \citenamefont {Ballance},
  \citenamefont {Thirumalai}, \citenamefont {Goodwin}, \citenamefont {Lucas},\
  and\ \citenamefont {Ballance}}]{stephenson2020:remote_entanglement}%
  \BibitemOpen
  \bibfield  {author} {\bibinfo {author} {\bibfnamefont {L.~J.}\ \bibnamefont
  {Stephenson}}, \bibinfo {author} {\bibfnamefont {D.~P.}\ \bibnamefont
  {Nadlinger}}, \bibinfo {author} {\bibfnamefont {B.~C.}\ \bibnamefont
  {Nichol}}, \bibinfo {author} {\bibfnamefont {S.}~\bibnamefont {An}}, \bibinfo
  {author} {\bibfnamefont {P.}~\bibnamefont {Drmota}}, \bibinfo {author}
  {\bibfnamefont {T.~G.}\ \bibnamefont {Ballance}}, \bibinfo {author}
  {\bibfnamefont {K.}~\bibnamefont {Thirumalai}}, \bibinfo {author}
  {\bibfnamefont {J.~F.}\ \bibnamefont {Goodwin}}, \bibinfo {author}
  {\bibfnamefont {D.~M.}\ \bibnamefont {Lucas}},\ and\ \bibinfo {author}
  {\bibfnamefont {C.~J.}\ \bibnamefont {Ballance}},\ }\bibfield  {title}
  {\bibinfo {title} {High-rate, high-fidelity entanglement of qubits across an
  elementary quantum network},\ }\href
  {https://doi.org/10.1103/PhysRevLett.124.110501} {\bibfield  {journal}
  {\bibinfo  {journal} {Phys. Rev. Lett.}\ }\textbf {\bibinfo {volume} {124}},\
  \bibinfo {pages} {110501} (\bibinfo {year} {2020})},\ \Eprint
  {https://arxiv.org/abs/arXiv:1911.10841} {arXiv:1911.10841} \BibitemShut
  {NoStop}%
\bibitem [{\citenamefont {Winzer}\ \emph {et~al.}(2018)\citenamefont {Winzer},
  \citenamefont {Neilson},\ and\ \citenamefont
  {Chraplyvy}}]{winzer2018:fiber-optic_transmission}%
  \BibitemOpen
  \bibfield  {author} {\bibinfo {author} {\bibfnamefont {P.~J.}\ \bibnamefont
  {Winzer}}, \bibinfo {author} {\bibfnamefont {D.~T.}\ \bibnamefont
  {Neilson}},\ and\ \bibinfo {author} {\bibfnamefont {A.~R.}\ \bibnamefont
  {Chraplyvy}},\ }\bibfield  {title} {\bibinfo {title} {Fiber-optic
  transmission and networking: the previous 20 and the next 20 years},\ }\href
  {https://doi.org/10.1364/OE.26.024190} {\bibfield  {journal} {\bibinfo
  {journal} {Opt. Express}\ }\textbf {\bibinfo {volume} {26}},\ \bibinfo
  {pages} {024190} (\bibinfo {year} {2018})}\BibitemShut {NoStop}%
\bibitem [{\citenamefont {Sibley}(2020)}]{sibley2020:optical_comm}%
  \BibitemOpen
  \bibfield  {author} {\bibinfo {author} {\bibfnamefont {M.}~\bibnamefont
  {Sibley}},\ }\href {https://doi.org/10.1007/978-3-030-34359-0} {\emph
  {\bibinfo {title} {Optical Communications}}},\ \bibinfo {edition} {3rd}\ ed.\
  (\bibinfo  {publisher} {Springer, Cham},\ \bibinfo {year} {2020})\BibitemShut
  {NoStop}%
\bibitem [{\citenamefont {McGuinness}\ \emph {et~al.}(2010)\citenamefont
  {McGuinness}, \citenamefont {Raymer}, \citenamefont {McKinstrie},\ and\
  \citenamefont {Radic}}]{mcguinness2010:freq_translation}%
  \BibitemOpen
  \bibfield  {author} {\bibinfo {author} {\bibfnamefont {H.~J.}\ \bibnamefont
  {McGuinness}}, \bibinfo {author} {\bibfnamefont {M.~G.}\ \bibnamefont
  {Raymer}}, \bibinfo {author} {\bibfnamefont {C.~J.}\ \bibnamefont
  {McKinstrie}},\ and\ \bibinfo {author} {\bibfnamefont {S.}~\bibnamefont
  {Radic}},\ }\bibfield  {title} {\bibinfo {title} {Quantum frequency
  translation of single-photon states in a photonic crystal fiber},\ }\href
  {https://doi.org/10.1103/PhysRevLett.105.093604} {\bibfield  {journal}
  {\bibinfo  {journal} {Phys. Rev. Lett.}\ }\textbf {\bibinfo {volume} {105}},\
  \bibinfo {pages} {093604} (\bibinfo {year} {2010})},\ \Eprint
  {https://arxiv.org/abs/arXiv:1006.4350} {arXiv:1006.4350} \BibitemShut
  {NoStop}%
\bibitem [{\citenamefont {Zaske}\ \emph {et~al.}(2012)\citenamefont {Zaske},
  \citenamefont {Lenhard}, \citenamefont {Ke\ss{}ler}, \citenamefont {Kettler},
  \citenamefont {Hepp}, \citenamefont {Arend}, \citenamefont {Albrecht},
  \citenamefont {Schulz}, \citenamefont {Jetter}, \citenamefont {Michler},\
  and\ \citenamefont {Becher}}]{zaske2012:freq_conversion}%
  \BibitemOpen
  \bibfield  {author} {\bibinfo {author} {\bibfnamefont {S.}~\bibnamefont
  {Zaske}}, \bibinfo {author} {\bibfnamefont {A.}~\bibnamefont {Lenhard}},
  \bibinfo {author} {\bibfnamefont {C.~A.}\ \bibnamefont {Ke\ss{}ler}},
  \bibinfo {author} {\bibfnamefont {J.}~\bibnamefont {Kettler}}, \bibinfo
  {author} {\bibfnamefont {C.}~\bibnamefont {Hepp}}, \bibinfo {author}
  {\bibfnamefont {C.}~\bibnamefont {Arend}}, \bibinfo {author} {\bibfnamefont
  {R.}~\bibnamefont {Albrecht}}, \bibinfo {author} {\bibfnamefont {W.-M.}\
  \bibnamefont {Schulz}}, \bibinfo {author} {\bibfnamefont {M.}~\bibnamefont
  {Jetter}}, \bibinfo {author} {\bibfnamefont {P.}~\bibnamefont {Michler}},\
  and\ \bibinfo {author} {\bibfnamefont {C.}~\bibnamefont {Becher}},\
  }\bibfield  {title} {\bibinfo {title} {Visible-to-telecom quantum frequency
  conversion of light from a single quantum emitter},\ }\href
  {https://doi.org/10.1103/PhysRevLett.109.147404} {\bibfield  {journal}
  {\bibinfo  {journal} {Phys. Rev. Lett.}\ }\textbf {\bibinfo {volume} {109}},\
  \bibinfo {pages} {147404} (\bibinfo {year} {2012})},\ \Eprint
  {https://arxiv.org/abs/arXiv:1204.6253} {arXiv:1204.6253} \BibitemShut
  {NoStop}%
\bibitem [{\citenamefont {Kim}\ \emph {et~al.}(2013)\citenamefont {Kim},
  \citenamefont {Clark},\ and\ \citenamefont
  {Gauthier}}]{kim2013:freq_conversion}%
  \BibitemOpen
  \bibfield  {author} {\bibinfo {author} {\bibfnamefont {J.}~\bibnamefont
  {Kim}}, \bibinfo {author} {\bibfnamefont {R.}~\bibnamefont {Clark}},\ and\
  \bibinfo {author} {\bibfnamefont {D.}~\bibnamefont {Gauthier}},\ }\bibfield
  {title} {\bibinfo {title} {Low-noise frequency downconversion for
  long-distance distribution of entangled atomic qubits},\ }in\ \href
  {https://doi.org/10.1109/PHOSST.2013.6614563} {\emph {\bibinfo {booktitle}
  {2013 IEEE Photonics Society Summer Topical Meeting Series}}}\ (\bibinfo
  {year} {2013})\ pp.\ \bibinfo {pages} {183--184}\BibitemShut {NoStop}%
\bibitem [{\citenamefont {Kasture}\ \emph {et~al.}(2016)\citenamefont
  {Kasture}, \citenamefont {Lenzini}, \citenamefont {Haylock}, \citenamefont
  {Boes}, \citenamefont {Mitchell}, \citenamefont {Streed},\ and\ \citenamefont
  {Lobino}}]{kasture2016:freq_conversion}%
  \BibitemOpen
  \bibfield  {author} {\bibinfo {author} {\bibfnamefont {S.}~\bibnamefont
  {Kasture}}, \bibinfo {author} {\bibfnamefont {F.}~\bibnamefont {Lenzini}},
  \bibinfo {author} {\bibfnamefont {B.}~\bibnamefont {Haylock}}, \bibinfo
  {author} {\bibfnamefont {A.}~\bibnamefont {Boes}}, \bibinfo {author}
  {\bibfnamefont {A.}~\bibnamefont {Mitchell}}, \bibinfo {author}
  {\bibfnamefont {E.~W.}\ \bibnamefont {Streed}},\ and\ \bibinfo {author}
  {\bibfnamefont {M.}~\bibnamefont {Lobino}},\ }\bibfield  {title} {\bibinfo
  {title} {Frequency conversion between {UV} and telecom wavelengths in a
  lithium niobate waveguide for quantum communication with {Yb${}^+$} trapped
  ions},\ }\href {https://doi.org/10.1088/2040-8978/18/10/104007} {\bibfield
  {journal} {\bibinfo  {journal} {J. Opt.}\ }\textbf {\bibinfo {volume} {18}},\
  \bibinfo {pages} {104007} (\bibinfo {year} {2016})},\ \Eprint
  {https://arxiv.org/abs/arXiv:1606.08127} {arXiv:1606.08127} \BibitemShut
  {NoStop}%
\bibitem [{\citenamefont {Kambs}\ \emph {et~al.}(2016)\citenamefont {Kambs},
  \citenamefont {Kettler}, \citenamefont {Bock}, \citenamefont {Becker},
  \citenamefont {Arend}, \citenamefont {Lenhard}, \citenamefont {Portalupi},
  \citenamefont {Jetter}, \citenamefont {Michler},\ and\ \citenamefont
  {Becher}}]{kambs2016:freq_conversion}%
  \BibitemOpen
  \bibfield  {author} {\bibinfo {author} {\bibfnamefont {B.}~\bibnamefont
  {Kambs}}, \bibinfo {author} {\bibfnamefont {J.}~\bibnamefont {Kettler}},
  \bibinfo {author} {\bibfnamefont {M.}~\bibnamefont {Bock}}, \bibinfo {author}
  {\bibfnamefont {J.~N.}\ \bibnamefont {Becker}}, \bibinfo {author}
  {\bibfnamefont {C.}~\bibnamefont {Arend}}, \bibinfo {author} {\bibfnamefont
  {A.}~\bibnamefont {Lenhard}}, \bibinfo {author} {\bibfnamefont {S.~L.}\
  \bibnamefont {Portalupi}}, \bibinfo {author} {\bibfnamefont {M.}~\bibnamefont
  {Jetter}}, \bibinfo {author} {\bibfnamefont {P.}~\bibnamefont {Michler}},\
  and\ \bibinfo {author} {\bibfnamefont {C.}~\bibnamefont {Becher}},\
  }\bibfield  {title} {\bibinfo {title} {Low-noise quantum frequency
  down-conversion of indistinguishable photons},\ }\href
  {https://doi.org/10.1364/OE.24.022250} {\bibfield  {journal} {\bibinfo
  {journal} {Opt. Express}\ }\textbf {\bibinfo {volume} {24}},\ \bibinfo
  {pages} {22250} (\bibinfo {year} {2016})}\BibitemShut {NoStop}%
\bibitem [{\citenamefont {Siverns}\ \emph {et~al.}(2017)\citenamefont
  {Siverns}, \citenamefont {Li},\ and\ \citenamefont
  {Quraishi}}]{siverns2017:freq_conversion}%
  \BibitemOpen
  \bibfield  {author} {\bibinfo {author} {\bibfnamefont {J.~D.}\ \bibnamefont
  {Siverns}}, \bibinfo {author} {\bibfnamefont {X.}~\bibnamefont {Li}},\ and\
  \bibinfo {author} {\bibfnamefont {Q.}~\bibnamefont {Quraishi}},\ }\bibfield
  {title} {\bibinfo {title} {Ion-photon entanglement and quantum frequency
  conversion with trapped {Ba$^+$} ions},\ }\href
  {https://doi.org/10.1364/AO.56.00B222} {\bibfield  {journal} {\bibinfo
  {journal} {Appl. Opt.}\ }\textbf {\bibinfo {volume} {56}},\ \bibinfo {pages}
  {B222} (\bibinfo {year} {2017})},\ \Eprint
  {https://arxiv.org/abs/arXiv:1701.02783} {arXiv:1701.02783} \BibitemShut
  {NoStop}%
\bibitem [{\citenamefont {R\"utz}\ \emph {et~al.}(2017)\citenamefont {R\"utz},
  \citenamefont {Luo}, \citenamefont {Suche},\ and\ \citenamefont
  {Silberhorn}}]{rutz2017:freq_conversion}%
  \BibitemOpen
  \bibfield  {author} {\bibinfo {author} {\bibfnamefont {H.}~\bibnamefont
  {R\"utz}}, \bibinfo {author} {\bibfnamefont {K.-H.}\ \bibnamefont {Luo}},
  \bibinfo {author} {\bibfnamefont {H.}~\bibnamefont {Suche}},\ and\ \bibinfo
  {author} {\bibfnamefont {C.}~\bibnamefont {Silberhorn}},\ }\bibfield  {title}
  {\bibinfo {title} {Quantum frequency conversion between infrared and
  ultraviolet},\ }\href {https://doi.org/10.1103/PhysRevApplied.7.024021}
  {\bibfield  {journal} {\bibinfo  {journal} {Phys. Rev. Applied}\ }\textbf
  {\bibinfo {volume} {7}},\ \bibinfo {pages} {024021} (\bibinfo {year}
  {2017})},\ \Eprint {https://arxiv.org/abs/arXiv:1610.03239}
  {arXiv:1610.03239} \BibitemShut {NoStop}%
\bibitem [{\citenamefont {Bock}\ \emph {et~al.}(2018)\citenamefont {Bock},
  \citenamefont {Eich}, \citenamefont {Kucera}, \citenamefont {Kreis},
  \citenamefont {Lenhard}, \citenamefont {Becher},\ and\ \citenamefont
  {Eschner}}]{bock2018:freq_conversion}%
  \BibitemOpen
  \bibfield  {author} {\bibinfo {author} {\bibfnamefont {M.}~\bibnamefont
  {Bock}}, \bibinfo {author} {\bibfnamefont {P.}~\bibnamefont {Eich}}, \bibinfo
  {author} {\bibfnamefont {S.}~\bibnamefont {Kucera}}, \bibinfo {author}
  {\bibfnamefont {M.}~\bibnamefont {Kreis}}, \bibinfo {author} {\bibfnamefont
  {A.}~\bibnamefont {Lenhard}}, \bibinfo {author} {\bibfnamefont
  {C.}~\bibnamefont {Becher}},\ and\ \bibinfo {author} {\bibfnamefont
  {J.}~\bibnamefont {Eschner}},\ }\bibfield  {title} {\bibinfo {title}
  {High-fidelity entanglement between a trapped ion and a telecom photon via
  quantum frequency conversion},\ }\href
  {https://doi.org/10.1038/s41467-018-04341-2} {\bibfield  {journal} {\bibinfo
  {journal} {Nat. Commun.}\ }\textbf {\bibinfo {volume} {9}},\ \bibinfo {pages}
  {1998} (\bibinfo {year} {2018})},\ \Eprint
  {https://arxiv.org/abs/arXiv:1710.04866} {arXiv:1710.04866} \BibitemShut
  {NoStop}%
\bibitem [{\citenamefont {Meraner}\ \emph {et~al.}(2020)\citenamefont
  {Meraner}, \citenamefont {Mazloom}, \citenamefont {Krutyanskiy},
  \citenamefont {Krcmarsky}, \citenamefont {Schupp}, \citenamefont {Fioretto},
  \citenamefont {Sekatski}, \citenamefont {Northup}, \citenamefont
  {Sangouard},\ and\ \citenamefont {Lanyon}}]{meraner2020:ion_network_node}%
  \BibitemOpen
  \bibfield  {author} {\bibinfo {author} {\bibfnamefont {M.}~\bibnamefont
  {Meraner}}, \bibinfo {author} {\bibfnamefont {A.}~\bibnamefont {Mazloom}},
  \bibinfo {author} {\bibfnamefont {V.}~\bibnamefont {Krutyanskiy}}, \bibinfo
  {author} {\bibfnamefont {V.}~\bibnamefont {Krcmarsky}}, \bibinfo {author}
  {\bibfnamefont {J.}~\bibnamefont {Schupp}}, \bibinfo {author} {\bibfnamefont
  {D.~A.}\ \bibnamefont {Fioretto}}, \bibinfo {author} {\bibfnamefont
  {P.}~\bibnamefont {Sekatski}}, \bibinfo {author} {\bibfnamefont {T.~E.}\
  \bibnamefont {Northup}}, \bibinfo {author} {\bibfnamefont {N.}~\bibnamefont
  {Sangouard}},\ and\ \bibinfo {author} {\bibfnamefont {B.~P.}\ \bibnamefont
  {Lanyon}},\ }\bibfield  {title} {\bibinfo {title} {Indistinguishable photons
  from a trapped-ion quantum network node},\ }\href
  {https://doi.org/10.1103/PhysRevA.102.052614} {\bibfield  {journal} {\bibinfo
   {journal} {Phys. Rev. A}\ }\textbf {\bibinfo {volume} {102}},\ \bibinfo
  {pages} {052614} (\bibinfo {year} {2020})},\ \Eprint
  {https://arxiv.org/abs/arXiv:1912.09259} {arXiv:1912.09259} \BibitemShut
  {NoStop}%
\bibitem [{\citenamefont {Hannegan}\ \emph {et~al.}(2021)\citenamefont
  {Hannegan}, \citenamefont {Saha}, \citenamefont {Siverns}, \citenamefont
  {Cassell}, \citenamefont {Waks},\ and\ \citenamefont
  {Quraishi}}]{hannegan2021:freq_conv}%
  \BibitemOpen
  \bibfield  {author} {\bibinfo {author} {\bibfnamefont {J.}~\bibnamefont
  {Hannegan}}, \bibinfo {author} {\bibfnamefont {U.}~\bibnamefont {Saha}},
  \bibinfo {author} {\bibfnamefont {J.~D.}\ \bibnamefont {Siverns}}, \bibinfo
  {author} {\bibfnamefont {J.}~\bibnamefont {Cassell}}, \bibinfo {author}
  {\bibfnamefont {E.}~\bibnamefont {Waks}},\ and\ \bibinfo {author}
  {\bibfnamefont {Q.}~\bibnamefont {Quraishi}},\ }\bibfield  {title} {\bibinfo
  {title} {C-band single photons from a trapped ion via two-stage frequency
  conversion},\ }\href {https://doi.org/10.1063/5.0059966} {\bibfield
  {journal} {\bibinfo  {journal} {Appl. Phys. Lett.}\ }\textbf {\bibinfo
  {volume} {119}},\ \bibinfo {pages} {084001} (\bibinfo {year} {2021})},\
  \Eprint {https://arxiv.org/abs/arXiv:2103.16450} {arXiv:2103.16450}
  \BibitemShut {NoStop}%
\bibitem [{\citenamefont {Safronova}\ and\ \citenamefont
  {Safronova}(2014)}]{safronova2014:laiii}%
  \BibitemOpen
  \bibfield  {author} {\bibinfo {author} {\bibfnamefont {U.~I.}\ \bibnamefont
  {Safronova}}\ and\ \bibinfo {author} {\bibfnamefont {M.~S.}\ \bibnamefont
  {Safronova}},\ }\bibfield  {title} {\bibinfo {title} {Relativistic many-body
  calculation of energies, transition rates, lifetimes, and multipole
  polarizabilities in {Cs}-like {La III}},\ }\href
  {https://doi.org/10.1103/PhysRevA.89.052515} {\bibfield  {journal} {\bibinfo
  {journal} {Phys. Rev. A}\ }\textbf {\bibinfo {volume} {89}},\ \bibinfo
  {pages} {052515} (\bibinfo {year} {2014})}\BibitemShut {NoStop}%
\bibitem [{\citenamefont {{de Laeter}}\ \emph {et~al.}(2003)\citenamefont {{de
  Laeter}}, \citenamefont {{B\"{o}hlke}}, \citenamefont {{De Bi\'{e}vre}},
  \citenamefont {Hidaka}, \citenamefont {Peiser}, \citenamefont {Rosman},\ and\
  \citenamefont {Taylor}}]{delaeter2003:atomic_weights}%
  \BibitemOpen
  \bibfield  {author} {\bibinfo {author} {\bibfnamefont {J.~R.}\ \bibnamefont
  {{de Laeter}}}, \bibinfo {author} {\bibfnamefont {J.~K.}\ \bibnamefont
  {{B\"{o}hlke}}}, \bibinfo {author} {\bibfnamefont {P.}~\bibnamefont {{De
  Bi\'{e}vre}}}, \bibinfo {author} {\bibfnamefont {H.}~\bibnamefont {Hidaka}},
  \bibinfo {author} {\bibfnamefont {H.~S.}\ \bibnamefont {Peiser}}, \bibinfo
  {author} {\bibfnamefont {K.~J.~R.}\ \bibnamefont {Rosman}},\ and\ \bibinfo
  {author} {\bibfnamefont {P.~D.~P.}\ \bibnamefont {Taylor}},\ }\bibfield
  {title} {\bibinfo {title} {Atomic weights of the elements: {R}eview 2000
  ({IUPAC} technical report)},\ }\href
  {https://doi.org/10.1351/pac200375060683} {\bibfield  {journal} {\bibinfo
  {journal} {Pure Appl. Chem.}\ }\textbf {\bibinfo {volume} {75}},\ \bibinfo
  {pages} {683} (\bibinfo {year} {2003})}\BibitemShut {NoStop}%
\bibitem [{\citenamefont {Olmschenk}\ \emph {et~al.}(2017)\citenamefont
  {Olmschenk}, \citenamefont {Banner}, \citenamefont {Hankes},\ and\
  \citenamefont {Nelson}}]{olmschenk2017:laiii_hyperfine}%
  \BibitemOpen
  \bibfield  {author} {\bibinfo {author} {\bibfnamefont {S.}~\bibnamefont
  {Olmschenk}}, \bibinfo {author} {\bibfnamefont {P.~R.}\ \bibnamefont
  {Banner}}, \bibinfo {author} {\bibfnamefont {J.}~\bibnamefont {Hankes}},\
  and\ \bibinfo {author} {\bibfnamefont {A.~M.}\ \bibnamefont {Nelson}},\
  }\bibfield  {title} {\bibinfo {title} {Optogalvanic spectroscopy of the
  hyperfine structure of the $5p^{6}5d$ ${}^2{D}_{3/2,5/2}$ and $5p^{6}4f$
  ${}^2{F}^o_{5/2,7/2}$ levels of {La III}},\ }\href
  {https://doi.org/10.1103/PhysRevA.96.032502} {\bibfield  {journal} {\bibinfo
  {journal} {Phys. Rev. A}\ }\textbf {\bibinfo {volume} {96}},\ \bibinfo
  {pages} {032502} (\bibinfo {year} {2017})},\ \Eprint
  {https://arxiv.org/abs/arXiv:1706.06015} {arXiv:1706.06015} \BibitemShut
  {NoStop}%
\bibitem [{\citenamefont {Li}\ \emph {et~al.}(2021)\citenamefont {Li},
  \citenamefont {Ma},\ and\ \citenamefont {Tang}}]{li2021:hyperfine_laiii}%
  \BibitemOpen
  \bibfield  {author} {\bibinfo {author} {\bibfnamefont {F.}~\bibnamefont
  {Li}}, \bibinfo {author} {\bibfnamefont {H.}~\bibnamefont {Ma}},\ and\
  \bibinfo {author} {\bibfnamefont {Y.-B.}\ \bibnamefont {Tang}},\ }\bibfield
  {title} {\bibinfo {title} {Relativistic coupled-cluster calculation of
  hyperfine-structure constants of {La}${}^{2+}$},\ }\href
  {https://doi.org/10.1088/1361-6455/abcdf0} {\bibfield  {journal} {\bibinfo
  {journal} {J. Phys. B: At. Mol. Opt. Phys.}\ }\textbf {\bibinfo {volume}
  {54}},\ \bibinfo {pages} {065003} (\bibinfo {year} {2021})}\BibitemShut
  {NoStop}%
\bibitem [{\citenamefont {Kramida}\ \emph {et~al.}(2020)\citenamefont
  {Kramida}, \citenamefont {{Yu. Ralchenko}}, \citenamefont {Reader},\ and\
  \citenamefont {{and NIST ASD Team}}}]{NIST:ASD_2020}%
  \BibitemOpen
  \bibfield  {author} {\bibinfo {author} {\bibfnamefont {A.}~\bibnamefont
  {Kramida}}, \bibinfo {author} {\bibnamefont {{Yu. Ralchenko}}}, \bibinfo
  {author} {\bibfnamefont {J.}~\bibnamefont {Reader}},\ and\ \bibinfo {author}
  {\bibnamefont {{and NIST ASD Team}}},\ }\href@noop {} {}\bibinfo
  {howpublished} {{NIST Atomic Spectra Database (ver. 5.8), [Online].
  Available: {\tt{https://physics.nist.gov/asd}} [2021, August 24]. National
  Institute of Standards and Technology, Gaithersburg, MD.}} (\bibinfo {year}
  {2020})\BibitemShut {NoStop}%
\bibitem [{\citenamefont {Urabe}\ \emph {et~al.}(1992)\citenamefont {Urabe},
  \citenamefont {Watanabe}, \citenamefont {Imajo},\ and\ \citenamefont
  {Hayasaka}}]{urabe1992:ca_ion_cooling}%
  \BibitemOpen
  \bibfield  {author} {\bibinfo {author} {\bibfnamefont {S.}~\bibnamefont
  {Urabe}}, \bibinfo {author} {\bibfnamefont {M.}~\bibnamefont {Watanabe}},
  \bibinfo {author} {\bibfnamefont {H.}~\bibnamefont {Imajo}},\ and\ \bibinfo
  {author} {\bibfnamefont {K.}~\bibnamefont {Hayasaka}},\ }\bibfield  {title}
  {\bibinfo {title} {Laser cooling of trapped {Ca}${}^+$ and measurement of the
  3 ${}^2{D}_{5/2}$ state lifetime},\ }\href
  {https://doi.org/10.1364/OL.17.001140} {\bibfield  {journal} {\bibinfo
  {journal} {Opt. Lett.}\ }\textbf {\bibinfo {volume} {17}},\ \bibinfo {pages}
  {1140} (\bibinfo {year} {1992})}\BibitemShut {NoStop}%
\bibitem [{\citenamefont {Madej}\ and\ \citenamefont
  {Sankey}(1990)}]{madej1990:sr_ion_laser_cooling}%
  \BibitemOpen
  \bibfield  {author} {\bibinfo {author} {\bibfnamefont {A.~A.}\ \bibnamefont
  {Madej}}\ and\ \bibinfo {author} {\bibfnamefont {J.~D.}\ \bibnamefont
  {Sankey}},\ }\bibfield  {title} {\bibinfo {title} {Single, trapped {Sr}${}^+$
  atom: laser cooling and quantum jumps by means of the $4d {}^2{D}_{5/2} - 5s
  {}^2{S}_{1/2}$ transition},\ }\href {https://doi.org/10.1364/OL.15.000634}
  {\bibfield  {journal} {\bibinfo  {journal} {Opt. Lett.}\ }\textbf {\bibinfo
  {volume} {15}},\ \bibinfo {pages} {634} (\bibinfo {year} {1990})}\BibitemShut
  {NoStop}%
\bibitem [{\citenamefont {Neuhauser}\ \emph {et~al.}(1978)\citenamefont
  {Neuhauser}, \citenamefont {Hohenstatt}, \citenamefont {Toschek},\ and\
  \citenamefont {Dehmelt}}]{neuhauser1978:ba_ion_cooling}%
  \BibitemOpen
  \bibfield  {author} {\bibinfo {author} {\bibfnamefont {W.}~\bibnamefont
  {Neuhauser}}, \bibinfo {author} {\bibfnamefont {M.}~\bibnamefont
  {Hohenstatt}}, \bibinfo {author} {\bibfnamefont {P.}~\bibnamefont
  {Toschek}},\ and\ \bibinfo {author} {\bibfnamefont {H.}~\bibnamefont
  {Dehmelt}},\ }\bibfield  {title} {\bibinfo {title} {Optical-sideband cooling
  of visible atom cloud confined in parabolic well},\ }\href
  {https://doi.org/10.1103/PhysRevLett.41.233} {\bibfield  {journal} {\bibinfo
  {journal} {Phys. Rev. Lett.}\ }\textbf {\bibinfo {volume} {41}},\ \bibinfo
  {pages} {233} (\bibinfo {year} {1978})}\BibitemShut {NoStop}%
\bibitem [{\citenamefont {Gray}\ \emph {et~al.}(1978)\citenamefont {Gray},
  \citenamefont {Whitley},\ and\ \citenamefont
  {Stroud}}]{gray1978:coherent_pop_trapping}%
  \BibitemOpen
  \bibfield  {author} {\bibinfo {author} {\bibfnamefont {H.~R.}\ \bibnamefont
  {Gray}}, \bibinfo {author} {\bibfnamefont {R.~M.}\ \bibnamefont {Whitley}},\
  and\ \bibinfo {author} {\bibfnamefont {C.~R.}\ \bibnamefont {Stroud}},\
  }\bibfield  {title} {\bibinfo {title} {Coherent trapping of atomic
  populations},\ }\href {https://doi.org/10.1364/OL.3.000218} {\bibfield
  {journal} {\bibinfo  {journal} {Opt. Lett.}\ }\textbf {\bibinfo {volume}
  {3}},\ \bibinfo {pages} {218} (\bibinfo {year} {1978})}\BibitemShut {NoStop}%
\bibitem [{\citenamefont {Bell}\ \emph {et~al.}(1991)\citenamefont {Bell},
  \citenamefont {Gill}, \citenamefont {Klein}, \citenamefont {Levick},
  \citenamefont {Tamm},\ and\ \citenamefont {Schnier}}]{bell:four-level}%
  \BibitemOpen
  \bibfield  {author} {\bibinfo {author} {\bibfnamefont {A.~S.}\ \bibnamefont
  {Bell}}, \bibinfo {author} {\bibfnamefont {P.}~\bibnamefont {Gill}}, \bibinfo
  {author} {\bibfnamefont {H.~A.}\ \bibnamefont {Klein}}, \bibinfo {author}
  {\bibfnamefont {A.~P.}\ \bibnamefont {Levick}}, \bibinfo {author}
  {\bibfnamefont {C.}~\bibnamefont {Tamm}},\ and\ \bibinfo {author}
  {\bibfnamefont {D.}~\bibnamefont {Schnier}},\ }\bibfield  {title} {\bibinfo
  {title} {Laser cooling of trapped ytterbium ions using a four-level
  optical-excitation scheme},\ }\href {https://doi.org/10.1103/PhysRevA.44.R20}
  {\bibfield  {journal} {\bibinfo  {journal} {Phys. Rev. A}\ }\textbf {\bibinfo
  {volume} {44}},\ \bibinfo {pages} {R20} (\bibinfo {year} {1991})}\BibitemShut
  {NoStop}%
\bibitem [{\citenamefont {Olmschenk}\ \emph {et~al.}(2007)\citenamefont
  {Olmschenk}, \citenamefont {Younge}, \citenamefont {Moehring}, \citenamefont
  {Matsukevich}, \citenamefont {Maunz},\ and\ \citenamefont
  {Monroe}}]{olmschenk:state-detect}%
  \BibitemOpen
  \bibfield  {author} {\bibinfo {author} {\bibfnamefont {S.}~\bibnamefont
  {Olmschenk}}, \bibinfo {author} {\bibfnamefont {K.~C.}\ \bibnamefont
  {Younge}}, \bibinfo {author} {\bibfnamefont {D.~L.}\ \bibnamefont
  {Moehring}}, \bibinfo {author} {\bibfnamefont {D.~N.}\ \bibnamefont
  {Matsukevich}}, \bibinfo {author} {\bibfnamefont {P.}~\bibnamefont {Maunz}},\
  and\ \bibinfo {author} {\bibfnamefont {C.}~\bibnamefont {Monroe}},\
  }\bibfield  {title} {\bibinfo {title} {Manipulation and detection of a
  trapped {Yb}${}^{+}$ hyperfine qubit},\ }\href
  {https://doi.org/10.1103/PhysRevA.76.052314} {\bibfield  {journal} {\bibinfo
  {journal} {Phys. Rev. A}\ }\textbf {\bibinfo {volume} {76}},\ \bibinfo
  {pages} {052314} (\bibinfo {year} {2007})},\ \Eprint
  {https://arxiv.org/abs/arXiv:0708.0657} {arXiv:0708.0657} \BibitemShut
  {NoStop}%
\bibitem [{\citenamefont {Ozeri}(2011)}]{ozeri2011:trapped_ion_tool_box}%
  \BibitemOpen
  \bibfield  {author} {\bibinfo {author} {\bibfnamefont {R.}~\bibnamefont
  {Ozeri}},\ }\bibfield  {title} {\bibinfo {title} {The trapped-ion qubit tool
  box},\ }\href {https://doi.org/10.1080/00107514.2011.603578} {\bibfield
  {journal} {\bibinfo  {journal} {Contemporary Physics}\ }\textbf {\bibinfo
  {volume} {52}},\ \bibinfo {pages} {531} (\bibinfo {year} {2011})},\ \Eprint
  {https://arxiv.org/abs/arXiv:1106.1190} {arXiv:1106.1190} \BibitemShut
  {NoStop}%
\bibitem [{\citenamefont {Benhelm}\ \emph {et~al.}(2008)\citenamefont
  {Benhelm}, \citenamefont {Kirchmair}, \citenamefont {Roos},\ and\
  \citenamefont {Blatt}}]{benhelm2008:ca43}%
  \BibitemOpen
  \bibfield  {author} {\bibinfo {author} {\bibfnamefont {J.}~\bibnamefont
  {Benhelm}}, \bibinfo {author} {\bibfnamefont {G.}~\bibnamefont {Kirchmair}},
  \bibinfo {author} {\bibfnamefont {C.~F.}\ \bibnamefont {Roos}},\ and\
  \bibinfo {author} {\bibfnamefont {R.}~\bibnamefont {Blatt}},\ }\bibfield
  {title} {\bibinfo {title} {Experimental quantum-information processing with
  ${}^{43}${Ca}${}^{+}$ ions},\ }\href
  {https://doi.org/10.1103/PhysRevA.77.062306} {\bibfield  {journal} {\bibinfo
  {journal} {Phys. Rev. A}\ }\textbf {\bibinfo {volume} {77}},\ \bibinfo
  {pages} {062306} (\bibinfo {year} {2008})},\ \Eprint
  {https://arxiv.org/abs/arXiv:0804.1261} {arXiv:0804.1261} \BibitemShut
  {NoStop}%
\bibitem [{\citenamefont {Arimondo}\ \emph {et~al.}(1977)\citenamefont
  {Arimondo}, \citenamefont {Inguscio},\ and\ \citenamefont
  {Violino}}]{arimondo1977:hyperfine}%
  \BibitemOpen
  \bibfield  {author} {\bibinfo {author} {\bibfnamefont {E.}~\bibnamefont
  {Arimondo}}, \bibinfo {author} {\bibfnamefont {M.}~\bibnamefont {Inguscio}},\
  and\ \bibinfo {author} {\bibfnamefont {P.}~\bibnamefont {Violino}},\
  }\bibfield  {title} {\bibinfo {title} {Experimental determinations of the
  hyperfine structure in the alkali atoms},\ }\href
  {https://doi.org/10.1103/RevModPhys.49.31} {\bibfield  {journal} {\bibinfo
  {journal} {Rev. Mod. Phys.}\ }\textbf {\bibinfo {volume} {49}},\ \bibinfo
  {pages} {31} (\bibinfo {year} {1977})}\BibitemShut {NoStop}%
\bibitem [{\citenamefont {Wineland}\ \emph {et~al.}(1998)\citenamefont
  {Wineland}, \citenamefont {Monroe}, \citenamefont {Itano}, \citenamefont
  {Leibfried}, \citenamefont {King},\ and\ \citenamefont
  {Meekhof}}]{wineland:nist-jnl}%
  \BibitemOpen
  \bibfield  {author} {\bibinfo {author} {\bibfnamefont {D.~J.}\ \bibnamefont
  {Wineland}}, \bibinfo {author} {\bibfnamefont {C.}~\bibnamefont {Monroe}},
  \bibinfo {author} {\bibfnamefont {W.~M.}\ \bibnamefont {Itano}}, \bibinfo
  {author} {\bibfnamefont {D.}~\bibnamefont {Leibfried}}, \bibinfo {author}
  {\bibfnamefont {B.~E.}\ \bibnamefont {King}},\ and\ \bibinfo {author}
  {\bibfnamefont {D.~M.}\ \bibnamefont {Meekhof}},\ }\bibfield  {title}
  {\bibinfo {title} {Experimental issues in coherent quantum-state manipulation
  of trapped atomic ions},\ }\href {https://doi.org/10.6028/jres.103.019}
  {\bibfield  {journal} {\bibinfo  {journal} {Journal of Research of the
  National Institute of Standards and Technology}\ }\textbf {\bibinfo {volume}
  {103}},\ \bibinfo {pages} {259} (\bibinfo {year} {1998})}\BibitemShut
  {NoStop}%
\bibitem [{\citenamefont {Acton}\ \emph {et~al.}(2006)\citenamefont {Acton},
  \citenamefont {Brickman}, \citenamefont {Haljan}, \citenamefont {Lee},
  \citenamefont {Deslauriers},\ and\ \citenamefont
  {Monroe}}]{acton2006:detection}%
  \BibitemOpen
  \bibfield  {author} {\bibinfo {author} {\bibfnamefont {M.}~\bibnamefont
  {Acton}}, \bibinfo {author} {\bibfnamefont {K.-A.}\ \bibnamefont {Brickman}},
  \bibinfo {author} {\bibfnamefont {P.~C.}\ \bibnamefont {Haljan}}, \bibinfo
  {author} {\bibfnamefont {P.~J.}\ \bibnamefont {Lee}}, \bibinfo {author}
  {\bibfnamefont {L.}~\bibnamefont {Deslauriers}},\ and\ \bibinfo {author}
  {\bibfnamefont {C.}~\bibnamefont {Monroe}},\ }\bibfield  {title} {\bibinfo
  {title} {Near-perfect simultaneous measurement of a qubit register},\ }\href
  {https://doi.org/10.26421/QIC6.6-1} {\bibfield  {journal} {\bibinfo
  {journal} {Quantum Information and Computation}\ }\textbf {\bibinfo {volume}
  {6}},\ \bibinfo {pages} {465} (\bibinfo {year} {2006})},\ \Eprint
  {https://arxiv.org/abs/arXiv:quant-ph/0511257} {arXiv:quant-ph/0511257}
  \BibitemShut {NoStop}%
\bibitem [{\citenamefont {Langer}(2006)}]{langer2006:thesis}%
  \BibitemOpen
  \bibfield  {author} {\bibinfo {author} {\bibfnamefont {C.~E.}\ \bibnamefont
  {Langer}},\ }\emph {\bibinfo {title} {High Fidelity Quantum Information
  Processing with Trapped Ions}},\ \href
  {https://www.proquest.com/docview/305351487} {Ph.D. thesis},\ \bibinfo
  {school} {University of Colorado at Boulder} (\bibinfo {year}
  {2006})\BibitemShut {NoStop}%
\bibitem [{\citenamefont {Noek}\ \emph {et~al.}(2013)\citenamefont {Noek},
  \citenamefont {Vrijsen}, \citenamefont {Gaultney}, \citenamefont {Mount},
  \citenamefont {Kim}, \citenamefont {Maunz},\ and\ \citenamefont
  {Kim}}]{noek2013:state_detect_yb}%
  \BibitemOpen
  \bibfield  {author} {\bibinfo {author} {\bibfnamefont {R.}~\bibnamefont
  {Noek}}, \bibinfo {author} {\bibfnamefont {G.}~\bibnamefont {Vrijsen}},
  \bibinfo {author} {\bibfnamefont {D.}~\bibnamefont {Gaultney}}, \bibinfo
  {author} {\bibfnamefont {E.}~\bibnamefont {Mount}}, \bibinfo {author}
  {\bibfnamefont {T.}~\bibnamefont {Kim}}, \bibinfo {author} {\bibfnamefont
  {P.}~\bibnamefont {Maunz}},\ and\ \bibinfo {author} {\bibfnamefont
  {J.}~\bibnamefont {Kim}},\ }\bibfield  {title} {\bibinfo {title} {High speed,
  high fidelity detection of an atomic hyperfine qubit},\ }\href
  {https://doi.org/10.1364/OL.38.004735} {\bibfield  {journal} {\bibinfo
  {journal} {Opt. Lett.}\ }\textbf {\bibinfo {volume} {38}},\ \bibinfo {pages}
  {4735} (\bibinfo {year} {2013})},\ \Eprint
  {https://arxiv.org/abs/arXiv:1304.3511} {arXiv:1304.3511} \BibitemShut
  {NoStop}%
\bibitem [{\citenamefont {Maiwald}\ \emph {et~al.}(2012)\citenamefont
  {Maiwald}, \citenamefont {Golla}, \citenamefont {Fischer}, \citenamefont
  {Bader}, \citenamefont {Heugel}, \citenamefont {Chalopin}, \citenamefont
  {Sondermann},\ and\ \citenamefont
  {Leuchs}}]{maiwald2012:collecting_photons_ion}%
  \BibitemOpen
  \bibfield  {author} {\bibinfo {author} {\bibfnamefont {R.}~\bibnamefont
  {Maiwald}}, \bibinfo {author} {\bibfnamefont {A.}~\bibnamefont {Golla}},
  \bibinfo {author} {\bibfnamefont {M.}~\bibnamefont {Fischer}}, \bibinfo
  {author} {\bibfnamefont {M.}~\bibnamefont {Bader}}, \bibinfo {author}
  {\bibfnamefont {S.}~\bibnamefont {Heugel}}, \bibinfo {author} {\bibfnamefont
  {B.}~\bibnamefont {Chalopin}}, \bibinfo {author} {\bibfnamefont
  {M.}~\bibnamefont {Sondermann}},\ and\ \bibinfo {author} {\bibfnamefont
  {G.}~\bibnamefont {Leuchs}},\ }\bibfield  {title} {\bibinfo {title}
  {Collecting more than half the fluorescence photons from a single ion},\
  }\href {https://doi.org/10.1103/PhysRevA.86.043431} {\bibfield  {journal}
  {\bibinfo  {journal} {Phys. Rev. A}\ }\textbf {\bibinfo {volume} {86}},\
  \bibinfo {pages} {043431} (\bibinfo {year} {2012})}\BibitemShut {NoStop}%
\bibitem [{\citenamefont {Chou}\ \emph {et~al.}(2017)\citenamefont {Chou},
  \citenamefont {Auchter}, \citenamefont {Lilieholm}, \citenamefont {Smith},\
  and\ \citenamefont {Blinov}}]{chou2017:parabolic}%
  \BibitemOpen
  \bibfield  {author} {\bibinfo {author} {\bibfnamefont {C.-K.}\ \bibnamefont
  {Chou}}, \bibinfo {author} {\bibfnamefont {C.}~\bibnamefont {Auchter}},
  \bibinfo {author} {\bibfnamefont {J.}~\bibnamefont {Lilieholm}}, \bibinfo
  {author} {\bibfnamefont {K.}~\bibnamefont {Smith}},\ and\ \bibinfo {author}
  {\bibfnamefont {B.}~\bibnamefont {Blinov}},\ }\bibfield  {title} {\bibinfo
  {title} {Note: Single ion imaging and fluorescence collection with a
  parabolic mirror trap},\ }\href {https://doi.org/10.1063/1.4996506}
  {\bibfield  {journal} {\bibinfo  {journal} {Rev. Sci. Inst.}\ }\textbf
  {\bibinfo {volume} {88}},\ \bibinfo {pages} {086101} (\bibinfo {year}
  {2017})},\ \Eprint {https://arxiv.org/abs/arXiv:1701.03187}
  {arXiv:1701.03187} \BibitemShut {NoStop}%
\bibitem [{\citenamefont {Ghadimi}\ \emph {et~al.}(2017)\citenamefont
  {Ghadimi}, \citenamefont {{Bl\={u}ms}}, \citenamefont {Norton}, \citenamefont
  {Fisher}, \citenamefont {Connell}, \citenamefont {Amini}, \citenamefont
  {Volin}, \citenamefont {Hayden}, \citenamefont {Pai}, \citenamefont
  {Kielpinski}, \citenamefont {Lobino},\ and\ \citenamefont
  {Streed}}]{ghadimi2017:diffractive_mirrors}%
  \BibitemOpen
  \bibfield  {author} {\bibinfo {author} {\bibfnamefont {M.}~\bibnamefont
  {Ghadimi}}, \bibinfo {author} {\bibfnamefont {V.}~\bibnamefont
  {{Bl\={u}ms}}}, \bibinfo {author} {\bibfnamefont {B.~G.}\ \bibnamefont
  {Norton}}, \bibinfo {author} {\bibfnamefont {P.~M.}\ \bibnamefont {Fisher}},
  \bibinfo {author} {\bibfnamefont {S.~C.}\ \bibnamefont {Connell}}, \bibinfo
  {author} {\bibfnamefont {J.~M.}\ \bibnamefont {Amini}}, \bibinfo {author}
  {\bibfnamefont {C.}~\bibnamefont {Volin}}, \bibinfo {author} {\bibfnamefont
  {H.}~\bibnamefont {Hayden}}, \bibinfo {author} {\bibfnamefont {C.-S.}\
  \bibnamefont {Pai}}, \bibinfo {author} {\bibfnamefont {D.}~\bibnamefont
  {Kielpinski}}, \bibinfo {author} {\bibfnamefont {M.}~\bibnamefont {Lobino}},\
  and\ \bibinfo {author} {\bibfnamefont {E.~W.}\ \bibnamefont {Streed}},\
  }\bibfield  {title} {\bibinfo {title} {Scalable ion-photon quantum interface
  based on integrated diffractive mirrors},\ }\href
  {https://doi.org/10.1038/s41534-017-0006-6} {\bibfield  {journal} {\bibinfo
  {journal} {npj Quantum Inf.}\ }\textbf {\bibinfo {volume} {3}},\ \bibinfo
  {pages} {4} (\bibinfo {year} {2017})},\ \Eprint
  {https://arxiv.org/abs/arXiv:1607.00100} {arXiv:1607.00100} \BibitemShut
  {NoStop}%
\bibitem [{\citenamefont {Crocker}\ \emph {et~al.}(2019)\citenamefont
  {Crocker}, \citenamefont {Lichtman}, \citenamefont {Sosnova}, \citenamefont
  {Carter}, \citenamefont {Scarano},\ and\ \citenamefont
  {Monroe}}]{crocker2019:pure_single_photons}%
  \BibitemOpen
  \bibfield  {author} {\bibinfo {author} {\bibfnamefont {C.}~\bibnamefont
  {Crocker}}, \bibinfo {author} {\bibfnamefont {M.}~\bibnamefont {Lichtman}},
  \bibinfo {author} {\bibfnamefont {K.}~\bibnamefont {Sosnova}}, \bibinfo
  {author} {\bibfnamefont {A.}~\bibnamefont {Carter}}, \bibinfo {author}
  {\bibfnamefont {S.}~\bibnamefont {Scarano}},\ and\ \bibinfo {author}
  {\bibfnamefont {C.}~\bibnamefont {Monroe}},\ }\bibfield  {title} {\bibinfo
  {title} {High purity single photons entangled with an atomic qubit},\ }\href
  {https://doi.org/10.1364/OE.27.028143} {\bibfield  {journal} {\bibinfo
  {journal} {Opt. Express}\ }\textbf {\bibinfo {volume} {27}},\ \bibinfo
  {pages} {28143} (\bibinfo {year} {2019})},\ \Eprint
  {https://arxiv.org/abs/arXiv:1812.01749} {arXiv:1812.01749} \BibitemShut
  {NoStop}%
\bibitem [{\citenamefont {Araneda}\ \emph {et~al.}(2020)\citenamefont
  {Araneda}, \citenamefont {Cerchiari}, \citenamefont {Higginbottom},
  \citenamefont {Holz}, \citenamefont {Lakhmanskiy}, \citenamefont
  {{Ob\u{s}il}}, \citenamefont {Colombe},\ and\ \citenamefont
  {Blatt}}]{araneda2020:panopticon_device}%
  \BibitemOpen
  \bibfield  {author} {\bibinfo {author} {\bibfnamefont {G.}~\bibnamefont
  {Araneda}}, \bibinfo {author} {\bibfnamefont {G.}~\bibnamefont {Cerchiari}},
  \bibinfo {author} {\bibfnamefont {D.~B.}\ \bibnamefont {Higginbottom}},
  \bibinfo {author} {\bibfnamefont {P.~C.}\ \bibnamefont {Holz}}, \bibinfo
  {author} {\bibfnamefont {K.}~\bibnamefont {Lakhmanskiy}}, \bibinfo {author}
  {\bibfnamefont {P.}~\bibnamefont {{Ob\u{s}il}}}, \bibinfo {author}
  {\bibfnamefont {Y.}~\bibnamefont {Colombe}},\ and\ \bibinfo {author}
  {\bibfnamefont {R.}~\bibnamefont {Blatt}},\ }\bibfield  {title} {\bibinfo
  {title} {The panopticon device: An integrated {Paul-trap-hemispherical}
  mirror system for quantum optics},\ }\href
  {https://doi.org/10.1063/5.0020661} {\bibfield  {journal} {\bibinfo
  {journal} {Rev. Sci. Inst.}\ }\textbf {\bibinfo {volume} {91}},\ \bibinfo
  {pages} {113201} (\bibinfo {year} {2020})},\ \Eprint
  {https://arxiv.org/abs/arXiv:2006.04828} {arXiv:2006.04828} \BibitemShut
  {NoStop}%
\bibitem [{\citenamefont {Walker}\ \emph {et~al.}(2020)\citenamefont {Walker},
  \citenamefont {Kashanian}, \citenamefont {Ward},\ and\ \citenamefont
  {Keller}}]{walker2020:photons_ion_cavity}%
  \BibitemOpen
  \bibfield  {author} {\bibinfo {author} {\bibfnamefont {T.}~\bibnamefont
  {Walker}}, \bibinfo {author} {\bibfnamefont {S.~V.}\ \bibnamefont
  {Kashanian}}, \bibinfo {author} {\bibfnamefont {T.}~\bibnamefont {Ward}},\
  and\ \bibinfo {author} {\bibfnamefont {M.}~\bibnamefont {Keller}},\
  }\bibfield  {title} {\bibinfo {title} {Improving the indistinguishability of
  single photons from an ion-cavity system},\ }\href
  {https://doi.org/10.1103/PhysRevA.102.032616} {\bibfield  {journal} {\bibinfo
   {journal} {Phys. Rev. A}\ }\textbf {\bibinfo {volume} {102}},\ \bibinfo
  {pages} {032616} (\bibinfo {year} {2020})},\ \Eprint
  {https://arxiv.org/abs/arXiv:1911.08442} {arXiv:1911.08442} \BibitemShut
  {NoStop}%
\bibitem [{\citenamefont {Takahashi}\ \emph {et~al.}(2020)\citenamefont
  {Takahashi}, \citenamefont {Kassa}, \citenamefont {Christoforou},\ and\
  \citenamefont {Keller}}]{takahashi2020:ion_cavity_coupling}%
  \BibitemOpen
  \bibfield  {author} {\bibinfo {author} {\bibfnamefont {H.}~\bibnamefont
  {Takahashi}}, \bibinfo {author} {\bibfnamefont {E.}~\bibnamefont {Kassa}},
  \bibinfo {author} {\bibfnamefont {C.}~\bibnamefont {Christoforou}},\ and\
  \bibinfo {author} {\bibfnamefont {M.}~\bibnamefont {Keller}},\ }\bibfield
  {title} {\bibinfo {title} {Strong coupling of a single ion to an optical
  cavity},\ }\href {https://doi.org/10.1103/PhysRevLett.124.013602} {\bibfield
  {journal} {\bibinfo  {journal} {Phys. Rev. Lett.}\ }\textbf {\bibinfo
  {volume} {124}},\ \bibinfo {pages} {013602} (\bibinfo {year} {2020})},\
  \Eprint {https://arxiv.org/abs/arXiv:1808.04031} {arXiv:1808.04031}
  \BibitemShut {NoStop}%
\bibitem [{\citenamefont {Schupp}\ \emph {et~al.}(2021)\citenamefont {Schupp},
  \citenamefont {Krcmarsky}, \citenamefont {Krutyanskiy}, \citenamefont
  {Meraner}, \citenamefont {Northup},\ and\ \citenamefont
  {Lanyon}}]{schupp2021:ion_photon_cavity}%
  \BibitemOpen
  \bibfield  {author} {\bibinfo {author} {\bibfnamefont {J.}~\bibnamefont
  {Schupp}}, \bibinfo {author} {\bibfnamefont {V.}~\bibnamefont {Krcmarsky}},
  \bibinfo {author} {\bibfnamefont {V.}~\bibnamefont {Krutyanskiy}}, \bibinfo
  {author} {\bibfnamefont {M.}~\bibnamefont {Meraner}}, \bibinfo {author}
  {\bibfnamefont {T.~E.}\ \bibnamefont {Northup}},\ and\ \bibinfo {author}
  {\bibfnamefont {B.~P.}~\bibnamefont {Lanyon}},\ }\bibfield  {title} {\bibinfo
  {title} {Interface between trapped-ion qubits and traveling photons with
  close-to-optimal efficiency},\ }\href
  {https://doi.org/10.1103/PRXQuantum.2.020331} {\bibfield  {journal} {\bibinfo
   {journal} {PRX Quantum}\ }\textbf {\bibinfo {volume} {2}},\ \bibinfo {pages}
  {020331} (\bibinfo {year} {2021})},\ \Eprint
  {https://arxiv.org/abs/arXiv:2105.02121} {arXiv:2105.02121} \BibitemShut
  {NoStop}%
\bibitem [{\citenamefont {Kobel}\ \emph {et~al.}(2021)\citenamefont {Kobel},
  \citenamefont {Breyer},\ and\ \citenamefont
  {{K\"{o}hl}}}]{kobel2021:ion_photon_cavity}%
  \BibitemOpen
  \bibfield  {author} {\bibinfo {author} {\bibfnamefont {P.}~\bibnamefont
  {Kobel}}, \bibinfo {author} {\bibfnamefont {M.}~\bibnamefont {Breyer}},\ and\
  \bibinfo {author} {\bibfnamefont {M.}~\bibnamefont {{K\"{o}hl}}},\ }\bibfield
   {title} {\bibinfo {title} {Deterministic spin-photon entanglement from a
  trapped ion in a fiber {Fabry-Perot} cavity},\ }\href
  {https://doi.org/10.1038/s41534-020-00338-2} {\bibfield  {journal} {\bibinfo
  {journal} {npj Quantum Inf.}\ }\textbf {\bibinfo {volume} {7}},\ \bibinfo
  {pages} {6} (\bibinfo {year} {2021})},\ \Eprint
  {https://arxiv.org/abs/arXiv:2005.09124} {arXiv:2005.09124} \BibitemShut
  {NoStop}%
\bibitem [{\citenamefont {Blinov}\ \emph {et~al.}(2004)\citenamefont {Blinov},
  \citenamefont {Moehring}, \citenamefont {Duan},\ and\ \citenamefont
  {Monroe}}]{blinov2004:ion-photon}%
  \BibitemOpen
  \bibfield  {author} {\bibinfo {author} {\bibfnamefont {B.~B.}\ \bibnamefont
  {Blinov}}, \bibinfo {author} {\bibfnamefont {D.~L.}\ \bibnamefont
  {Moehring}}, \bibinfo {author} {\bibfnamefont {L.-M.}\ \bibnamefont {Duan}},\
  and\ \bibinfo {author} {\bibfnamefont {C.}~\bibnamefont {Monroe}},\
  }\bibfield  {title} {\bibinfo {title} {Observation of entanglement between a
  single trapped atom and a single photon},\ }\href
  {https://doi.org/10.1038/nature02377} {\bibfield  {journal} {\bibinfo
  {journal} {Nature}\ }\textbf {\bibinfo {volume} {428}},\ \bibinfo {pages}
  {153} (\bibinfo {year} {2004})}\BibitemShut {NoStop}%
\bibitem [{\citenamefont {Moehring}\ \emph
  {et~al.}(2007{\natexlab{b}})\citenamefont {Moehring}, \citenamefont {Madsen},
  \citenamefont {Younge}, \citenamefont {{Kohn, Jr.}}, \citenamefont {Maunz},
  \citenamefont {Blinov}, \citenamefont {Duan},\ and\ \citenamefont
  {Monroe}}]{moehring2007:review}%
  \BibitemOpen
  \bibfield  {author} {\bibinfo {author} {\bibfnamefont {D.~L.}\ \bibnamefont
  {Moehring}}, \bibinfo {author} {\bibfnamefont {M.~J.}\ \bibnamefont
  {Madsen}}, \bibinfo {author} {\bibfnamefont {K.~C.}\ \bibnamefont {Younge}},
  \bibinfo {author} {\bibfnamefont {R.~N.}\ \bibnamefont {{Kohn, Jr.}}},
  \bibinfo {author} {\bibfnamefont {P.}~\bibnamefont {Maunz}}, \bibinfo
  {author} {\bibfnamefont {B.~B.}\ \bibnamefont {Blinov}}, \bibinfo {author}
  {\bibfnamefont {L.-M.}\ \bibnamefont {Duan}},\ and\ \bibinfo {author}
  {\bibfnamefont {C.}~\bibnamefont {Monroe}},\ }\bibfield  {title} {\bibinfo
  {title} {Quantum networking with photons and trapped atoms {(Invited)}},\
  }\href {https://doi.org/10.1364/JOSAB.24.000300} {\bibfield  {journal}
  {\bibinfo  {journal} {J. Opt. Soc. Am. B}\ }\textbf {\bibinfo {volume}
  {24}},\ \bibinfo {pages} {300} (\bibinfo {year}
  {2007}{\natexlab{b}})}\BibitemShut {NoStop}%
\bibitem [{\citenamefont {Luo}\ \emph {et~al.}(2009)\citenamefont {Luo},
  \citenamefont {Hayes}, \citenamefont {Manning}, \citenamefont {Matsukevich},
  \citenamefont {Maunz}, \citenamefont {Olmschenk}, \citenamefont {Sterk},\
  and\ \citenamefont {Monroe}}]{luo2009:protocols_q_network}%
  \BibitemOpen
  \bibfield  {author} {\bibinfo {author} {\bibfnamefont {L.}~\bibnamefont
  {Luo}}, \bibinfo {author} {\bibfnamefont {D.}~\bibnamefont {Hayes}}, \bibinfo
  {author} {\bibfnamefont {T.}~\bibnamefont {Manning}}, \bibinfo {author}
  {\bibfnamefont {D.}~\bibnamefont {Matsukevich}}, \bibinfo {author}
  {\bibfnamefont {P.}~\bibnamefont {Maunz}}, \bibinfo {author} {\bibfnamefont
  {S.}~\bibnamefont {Olmschenk}}, \bibinfo {author} {\bibfnamefont
  {J.}~\bibnamefont {Sterk}},\ and\ \bibinfo {author} {\bibfnamefont
  {C.}~\bibnamefont {Monroe}},\ }\bibfield  {title} {\bibinfo {title}
  {Protocols and techniques for a scalable atom-photon quantum network},\
  }\href {https://doi.org/10.1002/prop.200900093} {\bibfield  {journal}
  {\bibinfo  {journal} {Fortschritte der Physik}\ }\textbf {\bibinfo {volume}
  {57}},\ \bibinfo {pages} {1133} (\bibinfo {year} {2009})},\ \Eprint
  {https://arxiv.org/abs/arXiv:0906.1032} {arXiv:0906.1032} \BibitemShut
  {NoStop}%
\bibitem [{\citenamefont {Kim}\ \emph {et~al.}(2011)\citenamefont {Kim},
  \citenamefont {Maunz},\ and\ \citenamefont
  {Kim}}]{kim2011:collect_single_photons_ion}%
  \BibitemOpen
  \bibfield  {author} {\bibinfo {author} {\bibfnamefont {T.}~\bibnamefont
  {Kim}}, \bibinfo {author} {\bibfnamefont {P.}~\bibnamefont {Maunz}},\ and\
  \bibinfo {author} {\bibfnamefont {J.}~\bibnamefont {Kim}},\ }\bibfield
  {title} {\bibinfo {title} {Efficient collection of single photons emitted
  from a trapped ion into a single-mode fiber for scalable quantum-information
  processing},\ }\href {https://doi.org/10.1103/PhysRevA.84.063423} {\bibfield
  {journal} {\bibinfo  {journal} {Phys. Rev. A}\ }\textbf {\bibinfo {volume}
  {84}},\ \bibinfo {pages} {063423} (\bibinfo {year} {2011})},\ \Eprint
  {https://arxiv.org/abs/arXiv:1109.2268} {arXiv:1109.2268} \BibitemShut
  {NoStop}%
\bibitem [{\citenamefont {Duan}\ \emph {et~al.}(2006)\citenamefont {Duan},
  \citenamefont {Madsen}, \citenamefont {Moehring}, \citenamefont {Maunz},
  \citenamefont {{Kohn, Jr.}},\ and\ \citenamefont
  {Monroe}}]{duan2006:freq-qubit}%
  \BibitemOpen
  \bibfield  {author} {\bibinfo {author} {\bibfnamefont {L.-M.}\ \bibnamefont
  {Duan}}, \bibinfo {author} {\bibfnamefont {M.~J.}\ \bibnamefont {Madsen}},
  \bibinfo {author} {\bibfnamefont {D.~L.}\ \bibnamefont {Moehring}}, \bibinfo
  {author} {\bibfnamefont {P.}~\bibnamefont {Maunz}}, \bibinfo {author}
  {\bibfnamefont {R.~N.}\ \bibnamefont {{Kohn, Jr.}}},\ and\ \bibinfo {author}
  {\bibfnamefont {C.}~\bibnamefont {Monroe}},\ }\bibfield  {title} {\bibinfo
  {title} {Probabilistic quantum gates between remote atoms through
  interference of optical frequency qubits},\ }\href
  {https://doi.org/10.1103/PhysRevA.73.062324} {\bibfield  {journal} {\bibinfo
  {journal} {Phys. Rev. A}\ }\textbf {\bibinfo {volume} {73}},\ \bibinfo
  {pages} {062324} (\bibinfo {year} {2006})},\ \Eprint
  {https://arxiv.org/abs/arXiv:quant-ph/0603285} {arXiv:quant-ph/0603285}
  \BibitemShut {NoStop}%
\bibitem [{\citenamefont {Maunz}\ \emph {et~al.}(2009)\citenamefont {Maunz},
  \citenamefont {Olmschenk}, \citenamefont {Hayes}, \citenamefont
  {Matsukevich}, \citenamefont {Duan},\ and\ \citenamefont
  {Monroe}}]{maunz2009:heralded_gate}%
  \BibitemOpen
  \bibfield  {author} {\bibinfo {author} {\bibfnamefont {P.}~\bibnamefont
  {Maunz}}, \bibinfo {author} {\bibfnamefont {S.}~\bibnamefont {Olmschenk}},
  \bibinfo {author} {\bibfnamefont {D.}~\bibnamefont {Hayes}}, \bibinfo
  {author} {\bibfnamefont {D.~N.}\ \bibnamefont {Matsukevich}}, \bibinfo
  {author} {\bibfnamefont {L.-M.}\ \bibnamefont {Duan}},\ and\ \bibinfo
  {author} {\bibfnamefont {C.}~\bibnamefont {Monroe}},\ }\bibfield  {title}
  {\bibinfo {title} {Heralded quantum gate between remote quantum memories},\
  }\href {https://doi.org/10.1103/PhysRevLett.102.250502} {\bibfield  {journal}
  {\bibinfo  {journal} {Phys. Rev. Lett.}\ }\textbf {\bibinfo {volume} {102}},\
  \bibinfo {pages} {250502} (\bibinfo {year} {2009})},\ \Eprint
  {https://arxiv.org/abs/arXiv:0902.2136} {arXiv:0902.2136} \BibitemShut
  {NoStop}%
\bibitem [{\citenamefont {Brendel}\ \emph {et~al.}(1999)\citenamefont
  {Brendel}, \citenamefont {Gisin}, \citenamefont {Tittel},\ and\ \citenamefont
  {Zbinden}}]{brendel999:time-bin_qubit}%
  \BibitemOpen
  \bibfield  {author} {\bibinfo {author} {\bibfnamefont {J.}~\bibnamefont
  {Brendel}}, \bibinfo {author} {\bibfnamefont {N.}~\bibnamefont {Gisin}},
  \bibinfo {author} {\bibfnamefont {W.}~\bibnamefont {Tittel}},\ and\ \bibinfo
  {author} {\bibfnamefont {H.}~\bibnamefont {Zbinden}},\ }\bibfield  {title}
  {\bibinfo {title} {Pulsed energy-time entangled twin-photon source for
  quantum communication},\ }\href {https://doi.org/10.1103/PhysRevLett.82.2594}
  {\bibfield  {journal} {\bibinfo  {journal} {Phys. Rev. Lett.}\ }\textbf
  {\bibinfo {volume} {82}},\ \bibinfo {pages} {2594} (\bibinfo {year}
  {1999})},\ \Eprint {https://arxiv.org/abs/arXiv:quant-ph/9809034}
  {arXiv:quant-ph/9809034} \BibitemShut {NoStop}%
\bibitem [{\citenamefont {Barrett}\ and\ \citenamefont
  {Kok}(2005)}]{barrett2005:time-bin_qubit}%
  \BibitemOpen
  \bibfield  {author} {\bibinfo {author} {\bibfnamefont {S.~D.}\ \bibnamefont
  {Barrett}}\ and\ \bibinfo {author} {\bibfnamefont {P.}~\bibnamefont {Kok}},\
  }\bibfield  {title} {\bibinfo {title} {Efficient high-fidelity quantum
  computation using matter qubits and linear optics},\ }\href
  {https://doi.org/10.1103/PhysRevA.71.060310} {\bibfield  {journal} {\bibinfo
  {journal} {Phys. Rev. A}\ }\textbf {\bibinfo {volume} {71}},\ \bibinfo
  {pages} {060310(R)} (\bibinfo {year} {2005})},\ \Eprint
  {https://arxiv.org/abs/arXiv:quant-ph/0408040} {arXiv:quant-ph/0408040}
  \BibitemShut {NoStop}%
\bibitem [{\citenamefont {Bernien}\ \emph {et~al.}(2013)\citenamefont
  {Bernien}, \citenamefont {Hensen}, \citenamefont {Pfaff}, \citenamefont
  {Koolstra}, \citenamefont {Blok}, \citenamefont {Robledo}, \citenamefont
  {Taminiau}, \citenamefont {Markham}, \citenamefont {Twitchen}, \citenamefont
  {Childress},\ and\ \citenamefont
  {Hanson}}]{bernien2013:remote_nv_entanglement}%
  \BibitemOpen
  \bibfield  {author} {\bibinfo {author} {\bibfnamefont {H.}~\bibnamefont
  {Bernien}}, \bibinfo {author} {\bibfnamefont {B.}~\bibnamefont {Hensen}},
  \bibinfo {author} {\bibfnamefont {W.}~\bibnamefont {Pfaff}}, \bibinfo
  {author} {\bibfnamefont {G.}~\bibnamefont {Koolstra}}, \bibinfo {author}
  {\bibfnamefont {M.~S.}\ \bibnamefont {Blok}}, \bibinfo {author}
  {\bibfnamefont {L.}~\bibnamefont {Robledo}}, \bibinfo {author} {\bibfnamefont
  {T.~H.}\ \bibnamefont {Taminiau}}, \bibinfo {author} {\bibfnamefont
  {M.}~\bibnamefont {Markham}}, \bibinfo {author} {\bibfnamefont {D.~J.}\
  \bibnamefont {Twitchen}}, \bibinfo {author} {\bibfnamefont {L.}~\bibnamefont
  {Childress}},\ and\ \bibinfo {author} {\bibfnamefont {R.}~\bibnamefont
  {Hanson}},\ }\bibfield  {title} {\bibinfo {title} {Heralded entanglement
  between solid-state qubits separated by three metres},\ }\href
  {https://doi.org/10.1038/nature12016} {\bibfield  {journal} {\bibinfo
  {journal} {Nature}\ }\textbf {\bibinfo {volume} {497}},\ \bibinfo {pages}
  {86} (\bibinfo {year} {2013})},\ \Eprint
  {https://arxiv.org/abs/arXiv:1212.6136} {arXiv:1212.6136} \BibitemShut
  {NoStop}%
\bibitem [{fib()}]{fiber_attenuation_note}%
  \BibitemOpen
  \href@noop {} {}\bibinfo {note} {Based on SMF-28 Ultra optical fiber
  specificaton for max attenuation $\leq 0.32$ dB/km at the 1383 nm water
  absorption peak.}\BibitemShut {Stop}%
\bibitem [{smf(2019)}]{smf28ultra_datasheet}%
  \BibitemOpen
  \href
  {https://www.corning.com/media/worldwide/coc/documents/Fiber/PI-1424-AEN.pdf}
  {\emph {\bibinfo {title} {Corning\textsuperscript{\textregistered}
  SMF-28\textsuperscript{\textregistered} Ultra Optical Fiber}}},\ \bibinfo
  {organization} {Corning Incorporated} (\bibinfo {year} {2019})\BibitemShut
  {NoStop}%
\bibitem [{\citenamefont {Krutyanskiy}\ \emph {et~al.}(2019)\citenamefont
  {Krutyanskiy}, \citenamefont {Meraner}, \citenamefont {Schupp}, \citenamefont
  {Krcmarsky}, \citenamefont {Hainzer},\ and\ \citenamefont
  {Lanyon}}]{krutyanskiy2019:light-matter_50km}%
  \BibitemOpen
  \bibfield  {author} {\bibinfo {author} {\bibfnamefont {V.}~\bibnamefont
  {Krutyanskiy}}, \bibinfo {author} {\bibfnamefont {M.}~\bibnamefont
  {Meraner}}, \bibinfo {author} {\bibfnamefont {J.}~\bibnamefont {Schupp}},
  \bibinfo {author} {\bibfnamefont {V.}~\bibnamefont {Krcmarsky}}, \bibinfo
  {author} {\bibfnamefont {H.}~\bibnamefont {Hainzer}},\ and\ \bibinfo {author}
  {\bibfnamefont {B.~P.}\ \bibnamefont {Lanyon}},\ }\bibfield  {title}
  {\bibinfo {title} {Light-matter entanglement over 50 km of optical fibre},\
  }\href {https://doi.org/10.1038/s41534-019-0186-3} {\bibfield  {journal}
  {\bibinfo  {journal} {npj Quantum Inf.}\ }\textbf {\bibinfo {volume} {5}},\
  \bibinfo {pages} {72} (\bibinfo {year} {2019})},\ \Eprint
  {https://arxiv.org/abs/arXiv:1901.06317} {arXiv:1901.06317} \BibitemShut
  {NoStop}%
\bibitem [{rat()}]{rate_limit_note}%
  \BibitemOpen
  \href@noop {} {}\bibinfo {note} {This estimate of the distance-limited rate
  assumes $c$ as the speed of the quantum signal (photon) and $c/2$ as the
  speed of the classical detection signal, yielding a total time duration $t =
  L/c + L/(c/2) = 3L/c$. For a distance $L$ of about 5 km, $t = 50$ $\mu$s,
  which is already more than $10 \times$ the excited state
  lifetime.}\BibitemShut {Stop}%
\bibitem [{\citenamefont {Monroe}\ \emph {et~al.}(2014)\citenamefont {Monroe},
  \citenamefont {Raussendorf}, \citenamefont {Ruthven}, \citenamefont {Brown},
  \citenamefont {Maunz}, \citenamefont {Duan},\ and\ \citenamefont
  {Kim}}]{monroe2014:modular_qc}%
  \BibitemOpen
  \bibfield  {author} {\bibinfo {author} {\bibfnamefont {C.}~\bibnamefont
  {Monroe}}, \bibinfo {author} {\bibfnamefont {R.}~\bibnamefont {Raussendorf}},
  \bibinfo {author} {\bibfnamefont {A.}~\bibnamefont {Ruthven}}, \bibinfo
  {author} {\bibfnamefont {K.~R.}\ \bibnamefont {Brown}}, \bibinfo {author}
  {\bibfnamefont {P.}~\bibnamefont {Maunz}}, \bibinfo {author} {\bibfnamefont
  {L.-M.}\ \bibnamefont {Duan}},\ and\ \bibinfo {author} {\bibfnamefont
  {J.}~\bibnamefont {Kim}},\ }\bibfield  {title} {\bibinfo {title} {Large-scale
  modular quantum-computer architecture with atomic memory and photonic
  interconnects},\ }\href {https://doi.org/10.1103/PhysRevA.89.022317}
  {\bibfield  {journal} {\bibinfo  {journal} {Phys. Rev. A}\ }\textbf {\bibinfo
  {volume} {89}},\ \bibinfo {pages} {022317} (\bibinfo {year} {2014})},\
  \Eprint {https://arxiv.org/abs/arXiv:1208.0391} {arXiv:1208.0391}
  \BibitemShut {NoStop}%
\bibitem [{\citenamefont {Brown}\ \emph {et~al.}(2016)\citenamefont {Brown},
  \citenamefont {Kim},\ and\ \citenamefont
  {Monroe}}]{brown2016:co-design_tiqc}%
  \BibitemOpen
  \bibfield  {author} {\bibinfo {author} {\bibfnamefont {K.~R.}\ \bibnamefont
  {Brown}}, \bibinfo {author} {\bibfnamefont {J.}~\bibnamefont {Kim}},\ and\
  \bibinfo {author} {\bibfnamefont {C.}~\bibnamefont {Monroe}},\ }\bibfield
  {title} {\bibinfo {title} {Co-designing a scalable quantum computer with
  trapped atomic ions},\ }\href {https://doi.org/10.1038/npjqi.2016.34}
  {\bibfield  {journal} {\bibinfo  {journal} {npj Quantum Inf.}\ }\textbf
  {\bibinfo {volume} {2}},\ \bibinfo {pages} {16034} (\bibinfo {year}
  {2016})},\ \Eprint {https://arxiv.org/abs/arXiv:1602.02840}
  {arXiv:1602.02840} \BibitemShut {NoStop}%
\bibitem [{\citenamefont {Santra}\ \emph {et~al.}(2019)\citenamefont {Santra},
  \citenamefont {Muralidharan}, \citenamefont {Lichtman}, \citenamefont
  {Jiang}, \citenamefont {Monroe},\ and\ \citenamefont
  {Malinovsky}}]{santra2019:q_repeater_two_species}%
  \BibitemOpen
  \bibfield  {author} {\bibinfo {author} {\bibfnamefont {S.}~\bibnamefont
  {Santra}}, \bibinfo {author} {\bibfnamefont {S.}~\bibnamefont
  {Muralidharan}}, \bibinfo {author} {\bibfnamefont {M.}~\bibnamefont
  {Lichtman}}, \bibinfo {author} {\bibfnamefont {L.}~\bibnamefont {Jiang}},
  \bibinfo {author} {\bibfnamefont {C.}~\bibnamefont {Monroe}},\ and\ \bibinfo
  {author} {\bibfnamefont {V.~S.}\ \bibnamefont {Malinovsky}},\ }\bibfield
  {title} {\bibinfo {title} {Quantum repeaters based on two species trapped
  ions},\ }\href {https://doi.org/10.1088/1367-2630/ab2a45} {\bibfield
  {journal} {\bibinfo  {journal} {New J. Phys.}\ }\textbf {\bibinfo {volume}
  {21}},\ \bibinfo {pages} {073002} (\bibinfo {year} {2019})},\ \Eprint
  {https://arxiv.org/abs/arXiv:1811.10723} {arXiv:1811.10723} \BibitemShut
  {NoStop}%
\bibitem [{\citenamefont {Dhara}\ \emph {et~al.}(2021)\citenamefont {Dhara},
  \citenamefont {Linke}, \citenamefont {Waks}, \citenamefont {Guha},\ and\
  \citenamefont {Seshadreesan}}]{dhara2021:q_repeater}%
  \BibitemOpen
  \bibfield  {author} {\bibinfo {author} {\bibfnamefont {P.}~\bibnamefont
  {Dhara}}, \bibinfo {author} {\bibfnamefont {N.~M.}\ \bibnamefont {Linke}},
  \bibinfo {author} {\bibfnamefont {E.}~\bibnamefont {Waks}}, \bibinfo {author}
  {\bibfnamefont {S.}~\bibnamefont {Guha}},\ and\ \bibinfo {author}
  {\bibfnamefont {K.~P.}\ \bibnamefont {Seshadreesan}},\ }\bibfield  {title}
  {\bibinfo {title} {Multiplexed quantum repeaters based on dual-species
  trapped-ion systems},\ }\href@noop {} {\  (\bibinfo {year} {2021})},\ \Eprint
  {https://arxiv.org/abs/arXiv:2105.06707} {arXiv:2105.06707} \BibitemShut
  {NoStop}%
\bibitem [{\citenamefont {Kielpinski}\ \emph {et~al.}(2000)\citenamefont
  {Kielpinski}, \citenamefont {King}, \citenamefont {Myatt}, \citenamefont
  {Sackett}, \citenamefont {Turchette}, \citenamefont {Itano}, \citenamefont
  {Monroe}, \citenamefont {Wineland},\ and\ \citenamefont
  {Zurek}}]{kielpinski2000:sympathetic_cooling}%
  \BibitemOpen
  \bibfield  {author} {\bibinfo {author} {\bibfnamefont {D.}~\bibnamefont
  {Kielpinski}}, \bibinfo {author} {\bibfnamefont {B.~E.}\ \bibnamefont
  {King}}, \bibinfo {author} {\bibfnamefont {C.~J.}\ \bibnamefont {Myatt}},
  \bibinfo {author} {\bibfnamefont {C.~A.}\ \bibnamefont {Sackett}}, \bibinfo
  {author} {\bibfnamefont {Q.~A.}\ \bibnamefont {Turchette}}, \bibinfo {author}
  {\bibfnamefont {W.~M.}\ \bibnamefont {Itano}}, \bibinfo {author}
  {\bibfnamefont {C.}~\bibnamefont {Monroe}}, \bibinfo {author} {\bibfnamefont
  {D.~J.}\ \bibnamefont {Wineland}},\ and\ \bibinfo {author} {\bibfnamefont
  {W.~H.}\ \bibnamefont {Zurek}},\ }\bibfield  {title} {\bibinfo {title}
  {Sympathetic cooling of trapped ions for quantum logic},\ }\href
  {https://doi.org/10.1103/PhysRevA.61.032310} {\bibfield  {journal} {\bibinfo
  {journal} {Phys. Rev. A}\ }\textbf {\bibinfo {volume} {61}},\ \bibinfo
  {pages} {032310} (\bibinfo {year} {2000})},\ \Eprint
  {https://arxiv.org/abs/arXiv:quant-ph/9909035} {arXiv:quant-ph/9909035}
  \BibitemShut {NoStop}%
\bibitem [{\citenamefont {Home}\ \emph {et~al.}(2009)\citenamefont {Home},
  \citenamefont {Hanneke}, \citenamefont {Jost}, \citenamefont {Amini},
  \citenamefont {Leibfried},\ and\ \citenamefont
  {Wineland}}]{home2009:complete_methods_qip}%
  \BibitemOpen
  \bibfield  {author} {\bibinfo {author} {\bibfnamefont {J.~P.}\ \bibnamefont
  {Home}}, \bibinfo {author} {\bibfnamefont {D.}~\bibnamefont {Hanneke}},
  \bibinfo {author} {\bibfnamefont {J.~D.}\ \bibnamefont {Jost}}, \bibinfo
  {author} {\bibfnamefont {J.~M.}\ \bibnamefont {Amini}}, \bibinfo {author}
  {\bibfnamefont {D.}~\bibnamefont {Leibfried}},\ and\ \bibinfo {author}
  {\bibfnamefont {D.~J.}\ \bibnamefont {Wineland}},\ }\bibfield  {title}
  {\bibinfo {title} {Complete methods set for scalable ion trap quantum
  information processing},\ }\href {https://doi.org/10.1126/science.1177077}
  {\bibfield  {journal} {\bibinfo  {journal} {Science}\ }\textbf {\bibinfo
  {volume} {325}},\ \bibinfo {pages} {1227} (\bibinfo {year} {2009})},\ \Eprint
  {https://arxiv.org/abs/arXiv:0907.1865} {arXiv:0907.1865} \BibitemShut
  {NoStop}%
\bibitem [{\citenamefont {Lekitsch}\ \emph {et~al.}(2017)\citenamefont
  {Lekitsch}, \citenamefont {Weidt}, \citenamefont {Fowler}, \citenamefont
  {M{\o}lmer}, \citenamefont {Devitt}, \citenamefont {Wunderlich},\ and\
  \citenamefont {Hensinger}}]{lekitsch2017:blueprint_tiqc}%
  \BibitemOpen
  \bibfield  {author} {\bibinfo {author} {\bibfnamefont {B.}~\bibnamefont
  {Lekitsch}}, \bibinfo {author} {\bibfnamefont {S.}~\bibnamefont {Weidt}},
  \bibinfo {author} {\bibfnamefont {A.~G.}\ \bibnamefont {Fowler}}, \bibinfo
  {author} {\bibfnamefont {K.}~\bibnamefont {M{\o}lmer}}, \bibinfo {author}
  {\bibfnamefont {S.~J.}\ \bibnamefont {Devitt}}, \bibinfo {author}
  {\bibfnamefont {C.}~\bibnamefont {Wunderlich}},\ and\ \bibinfo {author}
  {\bibfnamefont {W.~K.}\ \bibnamefont {Hensinger}},\ }\bibfield  {title}
  {\bibinfo {title} {Blueprint for a microwave trapped ion quantum computer},\
  }\href {https://doi.org/10.1126/sciadv.1601540} {\bibfield  {journal}
  {\bibinfo  {journal} {Science Advances}\ }\textbf {\bibinfo {volume} {3}},\
  \bibinfo {pages} {e1601540} (\bibinfo {year} {2017})},\ \Eprint
  {https://arxiv.org/abs/arXiv:1508.00420} {arXiv:1508.00420} \BibitemShut
  {NoStop}%
\bibitem [{\citenamefont {Raghunandan}\ \emph {et~al.}(2020)\citenamefont
  {Raghunandan}, \citenamefont {Wolf}, \citenamefont {Ospelkaus}, \citenamefont
  {Schmidt},\ and\ \citenamefont
  {Weimer}}]{raghunandan2020:qsim_sympathetic_cooling}%
  \BibitemOpen
  \bibfield  {author} {\bibinfo {author} {\bibfnamefont {M.}~\bibnamefont
  {Raghunandan}}, \bibinfo {author} {\bibfnamefont {F.}~\bibnamefont {Wolf}},
  \bibinfo {author} {\bibfnamefont {C.}~\bibnamefont {Ospelkaus}}, \bibinfo
  {author} {\bibfnamefont {P.~O.}\ \bibnamefont {Schmidt}},\ and\ \bibinfo
  {author} {\bibfnamefont {H.}~\bibnamefont {Weimer}},\ }\bibfield  {title}
  {\bibinfo {title} {Initialization of quantum simulators by sympathetic
  cooling},\ }\href {https://doi.org/10.1126/sciadv.aaw9268} {\bibfield
  {journal} {\bibinfo  {journal} {Science Advances}\ }\textbf {\bibinfo
  {volume} {6}},\ \bibinfo {pages} {eaaw9268} (\bibinfo {year} {2020})},\
  \Eprint {https://arxiv.org/abs/arXiv:1901.02019} {arXiv:1901.02019}
  \BibitemShut {NoStop}%
\bibitem [{\citenamefont {Klimov}\ \emph {et~al.}(2003)\citenamefont {Klimov},
  \citenamefont {Guzm\'an}, \citenamefont {Retamal},\ and\ \citenamefont
  {Saavedra}}]{klimov2003:qutrit_ions}%
  \BibitemOpen
  \bibfield  {author} {\bibinfo {author} {\bibfnamefont {A.~B.}\ \bibnamefont
  {Klimov}}, \bibinfo {author} {\bibfnamefont {R.}~\bibnamefont {Guzm\'an}},
  \bibinfo {author} {\bibfnamefont {J.~C.}\ \bibnamefont {Retamal}},\ and\
  \bibinfo {author} {\bibfnamefont {C.}~\bibnamefont {Saavedra}},\ }\bibfield
  {title} {\bibinfo {title} {Qutrit quantum computer with trapped ions},\
  }\href {https://doi.org/10.1103/PhysRevA.67.062313} {\bibfield  {journal}
  {\bibinfo  {journal} {Phys. Rev. A}\ }\textbf {\bibinfo {volume} {67}},\
  \bibinfo {pages} {062313} (\bibinfo {year} {2003})}\BibitemShut {NoStop}%
\bibitem [{\citenamefont {Senko}\ \emph {et~al.}(2015)\citenamefont {Senko},
  \citenamefont {Richerme}, \citenamefont {Smith}, \citenamefont {Lee},
  \citenamefont {Cohen}, \citenamefont {Retzker},\ and\ \citenamefont
  {Monroe}}]{senko2015:int_spin_chain}%
  \BibitemOpen
  \bibfield  {author} {\bibinfo {author} {\bibfnamefont {C.}~\bibnamefont
  {Senko}}, \bibinfo {author} {\bibfnamefont {P.}~\bibnamefont {Richerme}},
  \bibinfo {author} {\bibfnamefont {J.}~\bibnamefont {Smith}}, \bibinfo
  {author} {\bibfnamefont {A.}~\bibnamefont {Lee}}, \bibinfo {author}
  {\bibfnamefont {I.}~\bibnamefont {Cohen}}, \bibinfo {author} {\bibfnamefont
  {A.}~\bibnamefont {Retzker}},\ and\ \bibinfo {author} {\bibfnamefont
  {C.}~\bibnamefont {Monroe}},\ }\bibfield  {title} {\bibinfo {title}
  {Realization of a quantum integer-spin chain with controllable
  interactions},\ }\href {https://doi.org/10.1103/PhysRevX.5.021026} {\bibfield
   {journal} {\bibinfo  {journal} {Phys. Rev. X}\ }\textbf {\bibinfo {volume}
  {5}},\ \bibinfo {pages} {021026} (\bibinfo {year} {2015})},\ \Eprint
  {https://arxiv.org/abs/arXiv:1410.0937} {arXiv:1410.0937} \BibitemShut
  {NoStop}%
\bibitem [{\citenamefont {Randall}\ \emph {et~al.}(2015)\citenamefont
  {Randall}, \citenamefont {Weidt}, \citenamefont {Standing}, \citenamefont
  {Lake}, \citenamefont {Webster}, \citenamefont {Murgia}, \citenamefont
  {Navickas}, \citenamefont {Roth},\ and\ \citenamefont
  {Hensinger}}]{randall2015:microwave_dressed_qubits_qutrits}%
  \BibitemOpen
  \bibfield  {author} {\bibinfo {author} {\bibfnamefont {J.}~\bibnamefont
  {Randall}}, \bibinfo {author} {\bibfnamefont {S.}~\bibnamefont {Weidt}},
  \bibinfo {author} {\bibfnamefont {E.~D.}\ \bibnamefont {Standing}}, \bibinfo
  {author} {\bibfnamefont {K.}~\bibnamefont {Lake}}, \bibinfo {author}
  {\bibfnamefont {S.~C.}\ \bibnamefont {Webster}}, \bibinfo {author}
  {\bibfnamefont {D.~F.}\ \bibnamefont {Murgia}}, \bibinfo {author}
  {\bibfnamefont {T.}~\bibnamefont {Navickas}}, \bibinfo {author}
  {\bibfnamefont {K.}~\bibnamefont {Roth}},\ and\ \bibinfo {author}
  {\bibfnamefont {W.~K.}\ \bibnamefont {Hensinger}},\ }\bibfield  {title}
  {\bibinfo {title} {Efficient preparation and detection of microwave
  dressed-state qubits and qutrits with trapped ions},\ }\href
  {https://doi.org/10.1103/PhysRevA.91.012322} {\bibfield  {journal} {\bibinfo
  {journal} {Phys. Rev. A}\ }\textbf {\bibinfo {volume} {91}},\ \bibinfo
  {pages} {012322} (\bibinfo {year} {2015})},\ \Eprint
  {https://arxiv.org/abs/arXiv:1409.1696} {arXiv:1409.1696} \BibitemShut
  {NoStop}%
\bibitem [{\citenamefont {Leupold}\ \emph {et~al.}(2018)\citenamefont
  {Leupold}, \citenamefont {Malinowski}, \citenamefont {Zhang}, \citenamefont
  {Negnevitsky}, \citenamefont {Cabello}, \citenamefont {Alonso},\ and\
  \citenamefont {Home}}]{leupold2018:q_contextual}%
  \BibitemOpen
  \bibfield  {author} {\bibinfo {author} {\bibfnamefont {F.~M.}\ \bibnamefont
  {Leupold}}, \bibinfo {author} {\bibfnamefont {M.}~\bibnamefont {Malinowski}},
  \bibinfo {author} {\bibfnamefont {C.}~\bibnamefont {Zhang}}, \bibinfo
  {author} {\bibfnamefont {V.}~\bibnamefont {Negnevitsky}}, \bibinfo {author}
  {\bibfnamefont {A.}~\bibnamefont {Cabello}}, \bibinfo {author} {\bibfnamefont
  {J.}~\bibnamefont {Alonso}},\ and\ \bibinfo {author} {\bibfnamefont {J.~P.}\
  \bibnamefont {Home}},\ }\bibfield  {title} {\bibinfo {title} {Sustained
  state-independent quantum contextual correlations from a single ion},\ }\href
  {https://doi.org/10.1103/PhysRevLett.120.180401} {\bibfield  {journal}
  {\bibinfo  {journal} {Phys. Rev. Lett.}\ }\textbf {\bibinfo {volume} {120}},\
  \bibinfo {pages} {180401} (\bibinfo {year} {2018})},\ \Eprint
  {https://arxiv.org/abs/arXiv:1706.07370} {arXiv:1706.07370} \BibitemShut
  {NoStop}%
\bibitem [{\citenamefont {Low}\ \emph {et~al.}(2020)\citenamefont {Low},
  \citenamefont {White}, \citenamefont {Cox}, \citenamefont {Day},\ and\
  \citenamefont {Senko}}]{low2020:qudit_ions}%
  \BibitemOpen
  \bibfield  {author} {\bibinfo {author} {\bibfnamefont {P.~J.}\ \bibnamefont
  {Low}}, \bibinfo {author} {\bibfnamefont {B.~M.}\ \bibnamefont {White}},
  \bibinfo {author} {\bibfnamefont {A.~A.}\ \bibnamefont {Cox}}, \bibinfo
  {author} {\bibfnamefont {M.~L.}\ \bibnamefont {Day}},\ and\ \bibinfo {author}
  {\bibfnamefont {C.}~\bibnamefont {Senko}},\ }\bibfield  {title} {\bibinfo
  {title} {Practical trapped-ion protocols for universal qudit-based quantum
  computing},\ }\href {https://doi.org/10.1103/PhysRevResearch.2.033128}
  {\bibfield  {journal} {\bibinfo  {journal} {Phys. Rev. Research}\ }\textbf
  {\bibinfo {volume} {2}},\ \bibinfo {pages} {033128} (\bibinfo {year}
  {2020})},\ \Eprint {https://arxiv.org/abs/arXiv:1907.08569}
  {arXiv:1907.08569} \BibitemShut {NoStop}%
\end{thebibliography}
\end{document}